\documentclass[prb,preprint,showpacs,showkeys,preprintnumbers,amsmath,amssymb]{revtex4}

\usepackage{graphicx}
\usepackage{bm}
\usepackage{subeqn}
\begin{document}
\title{Estimating  Intermittency   Exponent  in  Neutrally  Stratified
Atmospheric Surface Layer Flows: A Robust Framework based on Magnitude
Cumulant    and     Surrogate    Analyses}    \author{Sukanta    Basu}
\email{sukanta.basu@ttu.edu}  \affiliation{Department  of Geosciences,
Texas   Tech   University,  Lubbock,   TX   79409,  USA}   \author{Efi
Foufoula-Georgiou} \altaffiliation[Also  at ]{the National  Center for
Earth Surface  Dynamics.}  \affiliation{St. Anthony  Falls Laboratory,
University  of Minnesota,  Minneapolis, MN  55414,  USA} \author{Bruno
Lashermes}  \affiliation{St. Anthony  Falls Laboratory,  University of
Minnesota,  Minneapolis,  MN  55414, USA}  \author{Alain  Arn\'{e}odo}
\affiliation{Laboratorie de Physique,  Ecole Normale Sup\'{e}rieure de
Lyon, 46 All\'{e}e d'Italie, 69364 Lyon c\'{e}dex 07, France}

\date{\today}

\begin{abstract}
This study proposes a novel  framework based on magnitude cumulant and
surrogate analyses  to reliably detect and  estimate the intermittency
coefficient   from  short-length   coarse-resolution   turbulent  time
series. Intermittency  coefficients estimated  from a large  number of
neutrally stratified  atmospheric surface layer  turbulent series from
various field  campaigns are shown to remarkably concur  with well-known laboratory
experimental results.  In  addition, surrogate-based 
hypothesis  testing significantly reduces  the likelihood  of 
detecting a spurious non-zero intermittency coefficient from  non-intermittent series.
The discriminatory  power of the  proposed framework is  promising for
addressing the unresolved question of how atmospheric
stability affects the intermittency properties of boundary layer turbulence.
\end{abstract}



\maketitle
\medskip

\section{Introduction}  

Existence of small-scale intermittency  is an intriguing yet unsettled
topic in  contemporary turbulence research. Over a  number of decades,
researchers  had been  trying to  unravel intermittency  in turbulence
measurements  and  at the  same  time  formulating diverse  conceptual
models to  rationalize the observed intermittency.\cite{fris95,sche99}
Encouragingly,  `practical'  implications  of  intermittency  research
outcomes  are  also  being  appreciated by  the  numerical  turbulence
modeling community  and a critical knowledge transfer  is taking place
as            evidenced            by            the            recent
literature.\cite{scot99,velh01,basu04,anto05,burt05}

One of the most widely used statistics characterizing the intermittent
nature  of  turbulence   is  the  so-called  `intermittency  exponent'
($\mu$).\cite{moni75} From observational  data, $\mu$ can be estimated
directly  or indirectly  via  several methods.   The direct  estimates
typically  involve appropriate  characterization  of the  second-order
scaling   behavior   of  the   local   rate   of  energy   dissipation
($\varepsilon$) field.   In this respect,  several alternatives (e.g.,
second-order integral moment, two-point correlation function, spectral
density)   are   available   in  the   literature.\cite{mene91,sree93,pras97}
Recently, Cleve  et al.\cite{clev04} showed that  among various direct
approaches,       the       two-point       correlation       function
$\langle\varepsilon(x+r)\varepsilon(x)\rangle$     of    the    energy
dissipation field  provides the most  reliable estimates of  $\mu$. In
this case, one can write
\begin{equation}
\langle\varepsilon(x+r)\varepsilon(x)\rangle \sim r^{-\mu}
\end{equation}
where $r$  is within the  inertial range.  Here, the  angular brackets
denote spatial averaging.

The  direct  intermittency   exponent  estimation  methods  (based  on
$\varepsilon$) require very high-resolution (resolving on the order of
Kolmogorov  scale) data  series of  pristine quality.   Most commonly,
fast-response    hot   wire   measurements    are   used    for   this
purpose.\cite{sree93,pras97,clev04,anse84,cham84} However, acquisition
of hot wire data in a  natural setting could be quite challenging. For
example,  in  the  case  of  atmospheric boundary  layer  (ABL)  field
experiments, one  needs to perform meticulous hot  wire calibration at
short  regular intervals  in order  to account  for the  ever changing
(diurnally varying) ABL flow parameters.\cite{maru03,kunk06}

The  ABL community widely  uses sonic  anemometers for  turbulent flux
measurements  in fields.   In  contrast to  hot  wires, these  sensors
require   much   lesser    periodic   calibration   and   maintenance.
Unfortunately, path  lengths ($\sim$ 10 cm) and  sampling rate ($\sim$
20 Hz)  of conventional  sonic anemometers are  too coarse  for direct
$\mu$ estimation.  In this paper,  we will explore if sonic anemometer
measurements,  in lieu  of hot  wire data,  can be  reliably  used for
indirect estimation of $\mu$.

One  of  the  most   popular  indirect  $\mu$  estimation  methods  is
associated    with    the    scaling    of    sixth-order    structure
function.\cite{anse84,cham84,fris78}     With     certain    plausible
assumptions, one can show that:\cite{fris78}
\begin{equation}
S_6(r)  = \langle  \left(  u(x+r) -  u(x)\right)^6  \rangle =  \langle
\left(             \Delta            u\right)^6\rangle            \sim
r^2\langle\varepsilon(x)\varepsilon(x+r)\rangle
\end{equation}
Using Eqs.   1-2, one  gets: $\langle \left(  \Delta u\right)^6\rangle
\sim   r^{2-\mu}$.   Chambers   and  Antonia\cite{cham84}   used  this
relatively simple  indirect approach and  obtained $\mu \sim $  0.2 in
the atmospheric surface  layer.  From adequate statistical convergence
standpoint,  estimation of  higher-order (specifically  sixth-order in
this   context)   structure   functions   require   very   long   time
series.\cite{anse84}  For instance, Chambers  and Antonia\cite{cham84}
used several runs of 15 min  duration at a hot wire sampling frequency
of $\sim$ 1.2 kHz (i.e., $\sim$  1 million samples per series). On the
other  hand, 15-30  min sonic  anemometer-based ABL  turbulence series
would    typically    consist     of    only    20,000    to    40,000
samples. Understandably, the estimates  of $\mu$ from sonic anemometer
series using  the traditional sixth-order  structure function approach
will  not   be  very  reliable.   This  and  some   other  theoretical
considerations explained  in the next  section, motivate us to  use an
alternative  estimation   approach,  called  the   magnitude  cumulant
analysis, recently introduced by  Delour et al.\cite{delo01}.  In this
approach,   only  second-order   magnitude   cumulants  (rather   than
data-intensive sixth-order structure functions) are needed to estimate
the intermittency coefficient ($\mu$).

The inter-related objectives of this paper are twofold:
\begin{itemize}

\item[(1)] Assess the potential  of the magnitude cumulant analysis in
detecting    and    estimating    intermittency   from    short-length
coarse-resolution (sonic anemometer-acquired) ABL measurements; and

\item[(2)] Design a  rigorous hypothesis-testing framework which would
reduce the likelihood  of spurious detection of a  non-zero $\mu$ from
non-intermittent (monofractal) series.
\end{itemize}

The paper  is structured as  follows.  In Section  II, we  briefly
describe  the  magnitude cumulant  analysis  technique. The  Iterative
Amplitude Adjusted Fourier Transform (IAAFT) algorithm-based surrogate
analysis,  originally  developed by  the  chaos  theory community  for
detection of nonlinearity in  time series,\cite{schr96} is shown to be
very  robust and  reliable for  intermittency  hypothesis-testing.  The
IAAFT algorithm is presented  in Section III.  An extensive collection
of observational  data from  various field campaigns  is used  in this
study and Section IV provides brief descriptions of these field datasets.
Comprehensive  results of intermittency estimation are presented  in 
Section  V and  compared with
published literature wherever  possible. Lastly, Section VI summarizes
our  results and  discusses perspectives  for future  research  on the
intriguing question of how atmospheric stability affects intermittency
properties of boundary layer turbulence.

\section{Magnitude Cumulant Analysis}

In   the  turbulence  literature,   the  scaling   exponent  spectrum,
$\zeta_q$, is defined as:
\begin{equation}
S_q(r) = \langle (\Delta u)^q \rangle \sim r^{\zeta_q}
\end{equation}
where $S_q(r)$  is the so-called  $q$-th order structure  function. As
before, the  angular bracket  denotes spatial averaging  and $r$  is a
separation  distance that varies  within the  inertial range.   If the
scaling exponent  $\zeta_q$ is a  nonlinear function of $q$,  then the
field   is    called   `multifractal',   otherwise    it   is   termed
`monofractal'.\cite{fris95,bohr98}   In   the  traditional   structure
function approach, estimation of $\mu~(=2-\zeta_6)$ requires a log-log
plot of $S_6(r)$ vs $r$ and subsequent extraction of the slope using a
least-squares  linear  regression  fit  over  a  scaling  regime  (the
inertial range).   For short time  series, computation of  $S_6(r)$ is
problematic due to statistical  convergence. Moreover, this problem is
further compounded by the fact that even if the series is sufficiently
long  for   statistical  convergence  of   higher-order  moments,  the
underlying nature of intermittency might theoretically limit the range
over   which   the   equivalency   of  statistical   and   geometrical
interpretations of intermittency  hold.\cite{lash04} As a result, even
accurate estimates of  higher-order statistical moment will degenerate
to  a linear  behavior of  $\zeta_q$ for  $q$ larger than some $q_{max}$ prohibiting
therefore an accurate estimation  of intermittency using the structure
function approach (see Lashermes et al.\cite{lash04} for details).  An
alternative   reliable   method,   first   advocated  by   Delour   et
al.\cite{delo01}, is to use  the magnitude cumulant analysis.  In this
approach, the relationship between  the moments of velocity increments
($\Delta   u$)    and   the   magnitude    cumulants   ($C_n$)   reads
as:\cite{delo01,chev06}
\begin{equation}
\langle    |\Delta    u    |^q\rangle    =    \exp(\sum_{n=1}^{\infty}
C_n(r)\frac{q^n}{n!})
\end{equation}  
where
\begin{subequations}
\begin{equation}
C_1(r) \equiv \langle \ln | \Delta u| \rangle \sim -c_1\ln(r)
\end{equation}
\begin{equation}
C_2(r) \equiv  \langle (\ln | \Delta u  | )^2 \rangle -  \langle \ln |
\Delta u | \rangle^2 \sim -c_2\ln(r)
\end{equation}
\begin{equation}
\begin{array}{l}
C_3(r) \equiv \langle  (\ln | \Delta u |)^3 \rangle  - 3\langle (\ln |
\Delta  u  |)^2   \rangle  \langle  \ln  |  \Delta   u  |  \rangle  \\
\hspace{0.5in} + 2\langle \ln |\Delta u | \rangle^3 \sim -c_3\ln(r)
\end{array}
\end{equation}
\end{subequations}
From Eqs. 3-5,  it is straightforward to express  the scaling exponent
spectrum as:\cite{delo01}
\begin{equation}
\zeta_q = -\sum_{n=1}^{\infty} c_n\frac{q^n}{n!}
\end{equation}
Furthermore,  by invoking a  relationship between  velocity increments
$(\Delta  u)$ and  local  rate of  dissipation fields  $(\varepsilon)$
(similar to Eq.  2), one can arrive at:\cite{male00}
\begin{equation}
\mu \simeq 9c_2
\end{equation}
Therefore, estimation  of the intermittency exponent  $\mu$ would only
require the computation of  second-order magnitude cumulant, i.e., the
second central moment of $ln|\Delta u|$ (Eq. 5b).

For large separation ($r \to  L_i$, where $L_i$ is the integral length
scale),  it  is  well  documented that  the  probability  distribution
function  (pdf)  of  velocity  increments ($\Delta  u$)  approaches  a
Gaussian distribution. For this scenario, the following results can be
derived analytically:
\begin{subequations}
\begin{eqnarray}
C_1(r) \to \frac{1}{2}(-\gamma -\ln(2))  = -0.64 \\ C_2(r) \to \pi^2/8
= 1.23\\ C_3(r) \to -\frac{7}{4}\zeta(3) = -2.1
\end{eqnarray}
\end{subequations}
where $\gamma$ is the Euler  Gamma constant = 0.577216, and $\zeta(3)$
is  Ap\'{e}ry's  constant  =  1.2020569. These  asymptotic  values  of
$C_1(r)$,  $C_2(r)$, and $C_3(r)$  would be  very useful  to demarcate
scaling regions in the case of short-length time series.

It is noted that instead  of a physical space-based magnitude cumulant
analysis approach  (i.e, Eqs. 4-5),  one could also  use wavelet-based
magnitude      cumulant      analysis      (see      Venugopal      et
al.\cite{venu06a,venu06b}   for   a   geophysical   application).    A
wavelet-based approach becomes necessary for nonstationary signals and
signals  with  H\"{o}lder  exponents   ($h$)  outside  the  window  of
$[0~1]$.\cite{muzy93,arne95}  In turbulence,  $\langle  h \rangle$  is
close to K41 value of $1/3$  and to best of our knowledge always found
to be within the window of $h \in [0~1]$.\cite{muzy91,verg93} Thus, in
the present study we  decided to employ physical space-based magnitude
cumulant analysis approach.

Magnitude cumulant analysis of  a synthetic fractional Brownian motion
with $h = 1/3$ (which  displays K41 like $k^{-5/3}$ spectrum) is shown
in Fig.  1.  The dashed  line in $C_1(r)$  vs.  $\ln(r)$ plot  has the
expected slope  of $1/3$.  For  almost the entire scaling  range, both
$C_2(r)$ and $C_3(r)$ remain  close to the theoretical Gaussian values
of $1.23$ and $-2.1$, respectively. This signal does not show any sign
of multifractality  (expected) as the  slope of $C_2(r)$  vs. $\ln(r)$
cannot be claimed to be different from zero.

\begin{widetext}
\begin{figure*}[ht]
\includegraphics[width=2.5in]{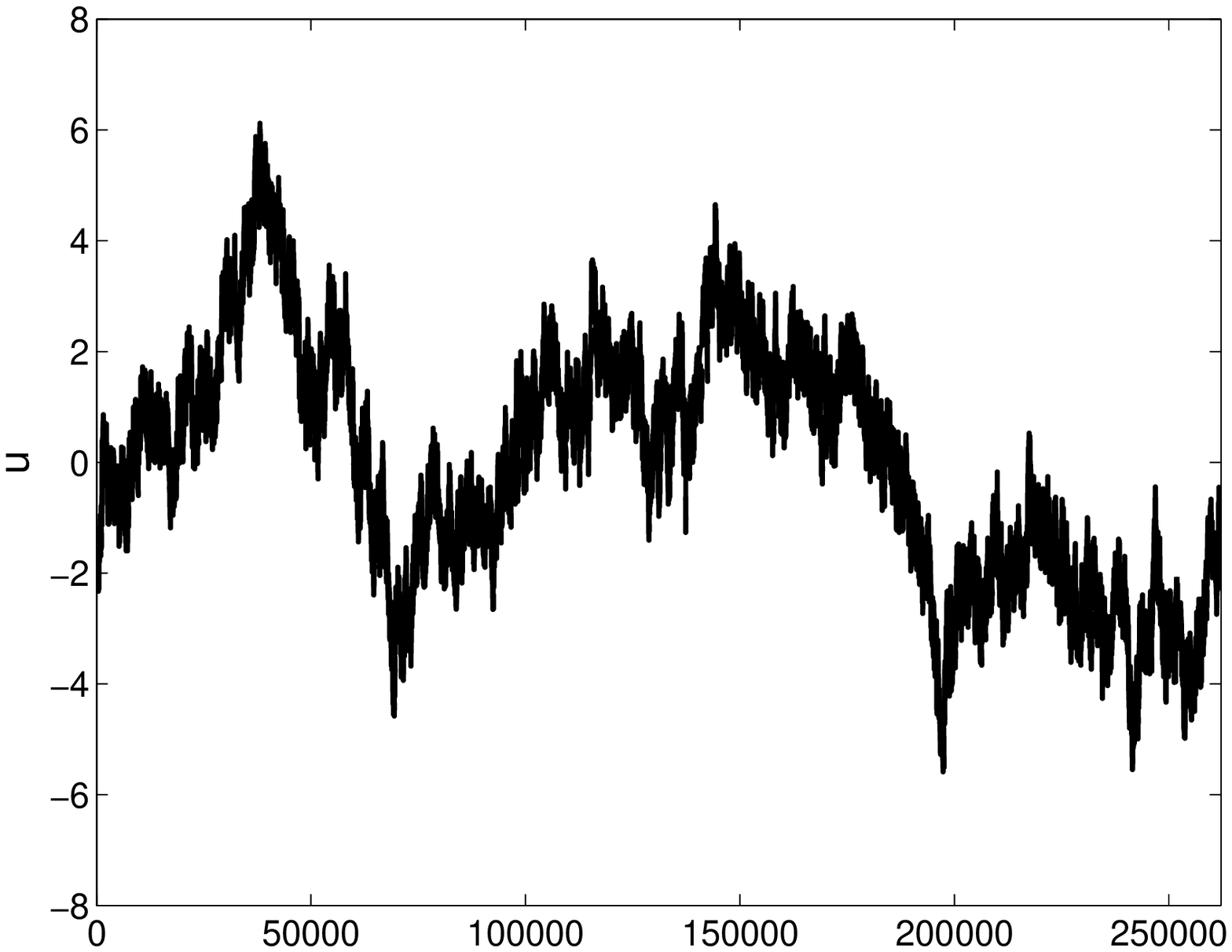}
\includegraphics[width=2.5in]{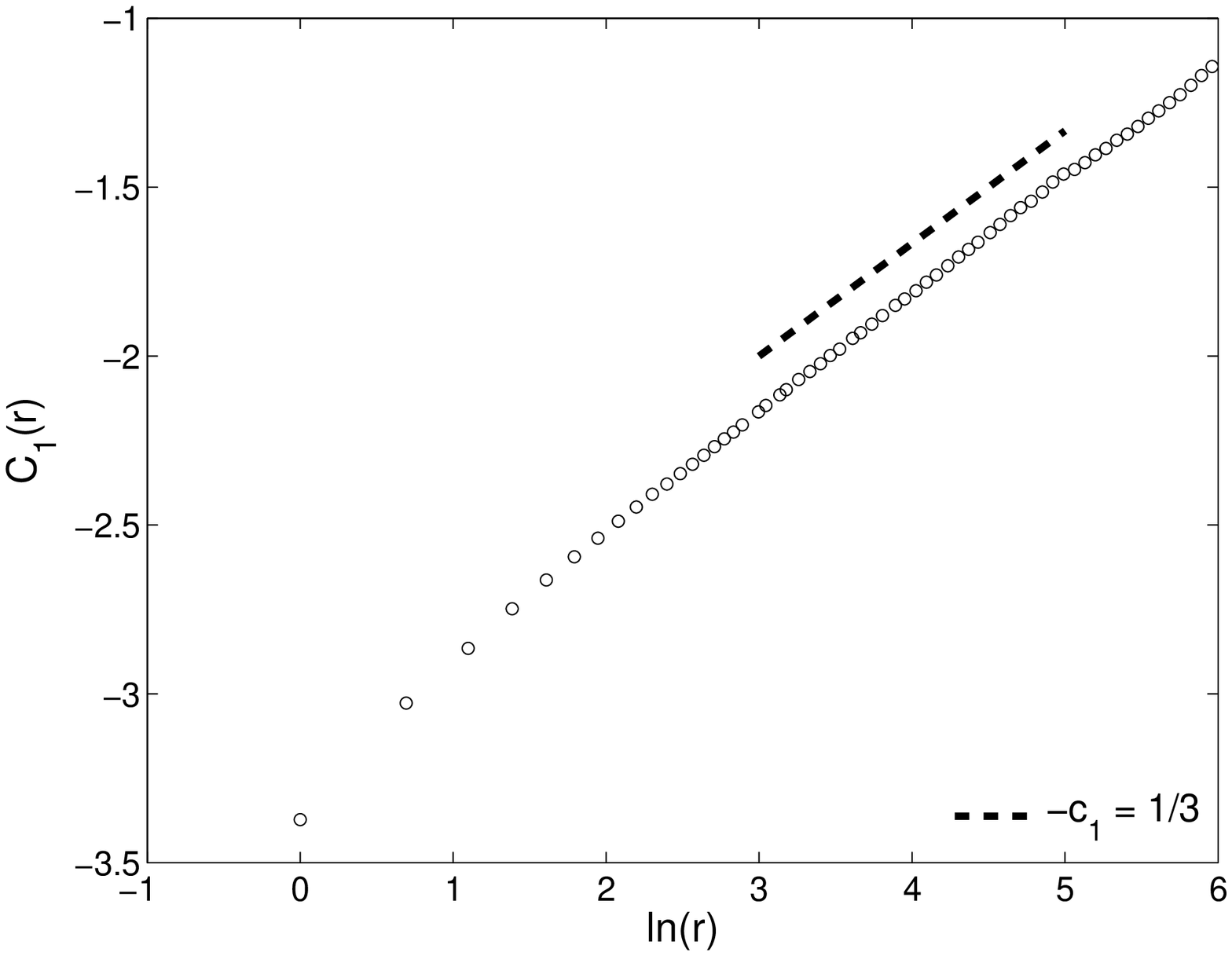}\\
\includegraphics[width=2.5in]{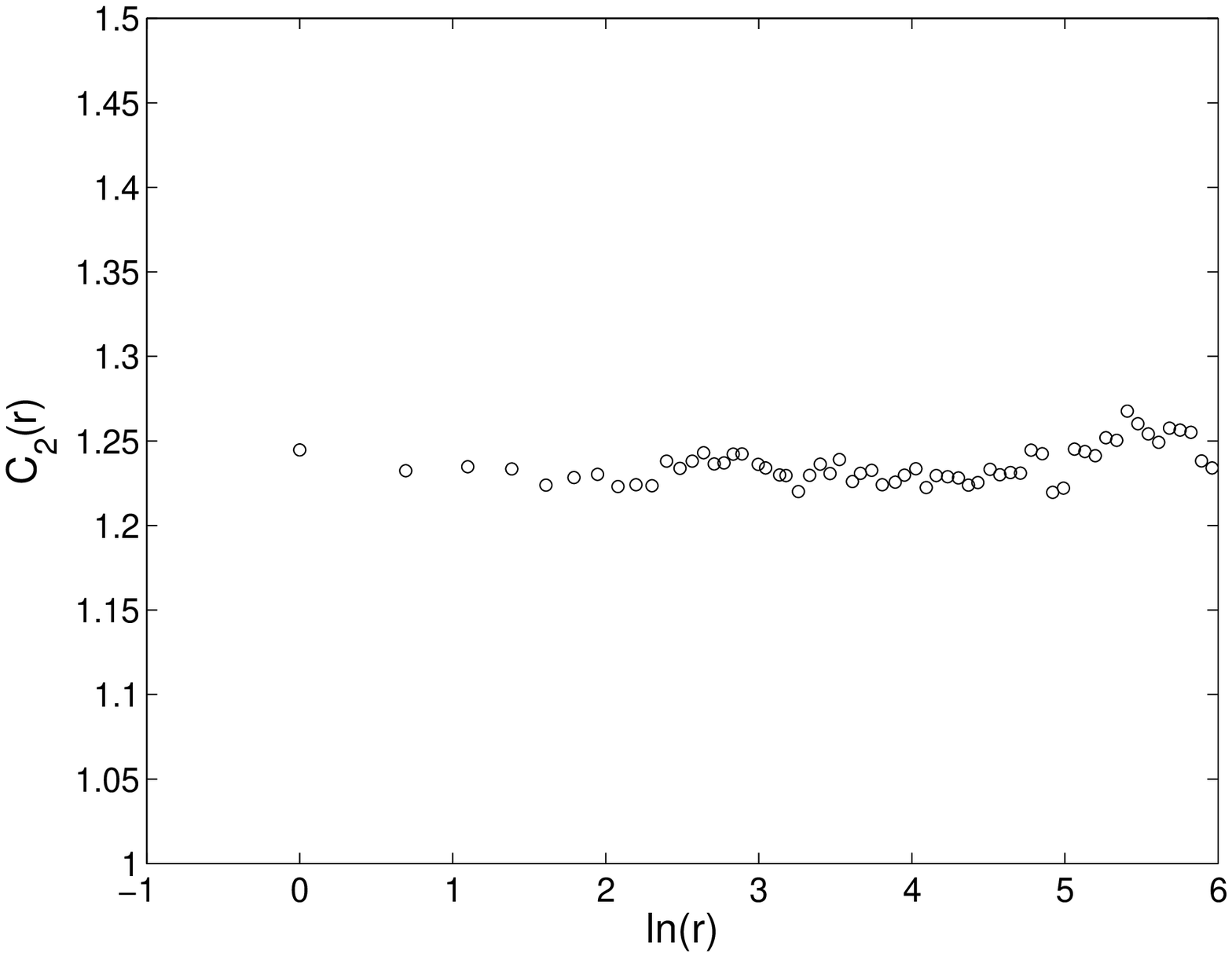}
\includegraphics[width=2.5in]{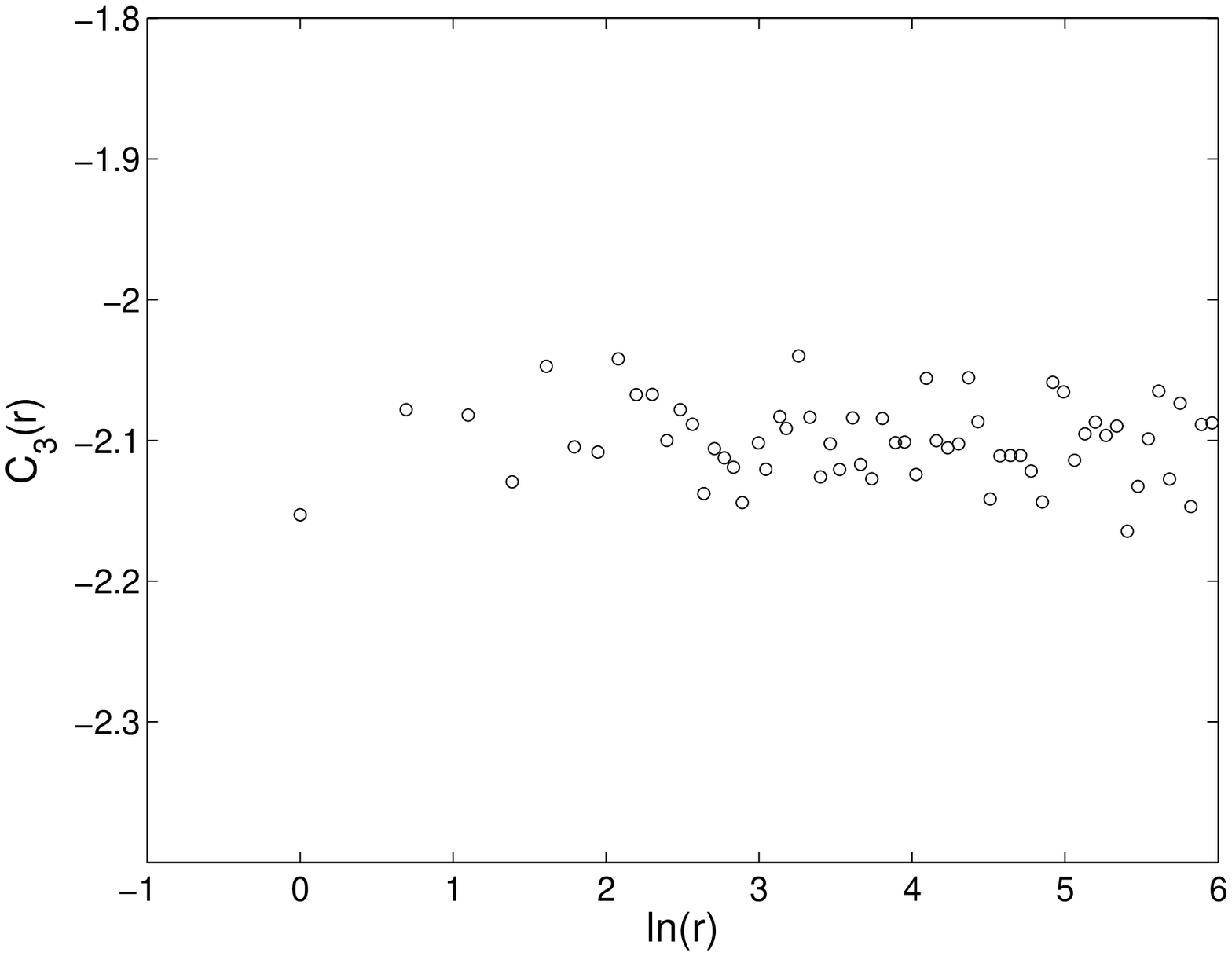}\\
\caption{A   synthetic  fractional   Brownian  motion   $(h   =  1/3)$
(top-left).   The magnitude  cumulants $C_1(r)$  (top-right), $C_2(r)$
(bottom-left), and $C_3(r)$ (bottom-right) are also shown. The dashed
line in the top-right subplot shows the slope $ -c_1 = 1/3$.}
\end{figure*}
\end{widetext}

\section{Surrogate Analysis}

Noise is omnipresent in any measured signal and turbulence signals are
no exceptions.  In  addition to noise, limited amount  of data (finite
sample   settings)  in   most  field   measurements   could  challenge
intermittency  detection  and   estimation  even  with  the  magnitude
cumulant  analysis method  (e.g., assessment  of  a  small non-zero
slope in the $C_2(r)$ vs. $ln(r)$ plots).  In this paper, we utilize a
hypothesis-testing   framework,  based on  surrogate   analysis,  in
conjunction  with  magnitude   cumulant  analysis,  for  detecting  and
accurately estimating   intermittency    from   short-length   sonic   
anemometer measurements.

The concept of surrogates (stochastic realizations which preserve only
certain characteristics  of a process)  was introduced into  the chaos
theory literature to provide a  rigorous statistical test for the null
hypothesis that an observed time series has been generated by a linear
stochastic  process  (see   Theiler  et  al.,\cite{thei92}  Kantz  and
Schreiber,\cite{kant97}  Basu and  Foufoula-Georgiou,\cite{basu02} and
the  references  there  in).  Over  the years,  several  varieties  of
surrogates  (randomly shuffled  surrogates,  Fourier phase  randomized
surrogates,  iterative amplitude  adjusted Fourier  transform  - IAAFT
surrogates, stochastic IAAFT surrogates, and so on) have been proposed
in the literature.\cite{vene06a} In this  paper, we will use the IAAFT
algorithm  proposed  by   Schreiber  and  Schmitz.\cite{schr96}  IAAFT
surrogates  preserve the  correlation structure  (thus  power spectrum
owing to Wiener-Khinchin theorem) and the probability density function
of    a    given     time    series.     Apart    from    nonlinearity
detection,\cite{schr96,basu02}  the IAAFT  surrogates  have also  been
used to  define a precipitation  forecast quality index,\cite{venu05}
and to generate synthetic cloud fields.\cite{vene06b}

In    the     turbulence    literature,    surrogate    analysis-based
hypothesis-testing is  virtually nonexistent with an  exception of the
paper  by Nikora et  al.\cite{niko01} They  used simple  Fourier phase
randomization approach (pdf of  the original turbulence series was not
preserved) in identifying the  effects of turbulence intermittency and
spectral energy  flux. In comparison  to the Fourier  phase randomized
approach,  the  IAAFT algorithm  used  in  the  present study  designs
stronger  statistical  test (owing  to  its  ability  to preserve  the
integral pdf of  the original signal) for the  null hypothesis that an
observed turbulence series is non-intermittent.

In  Fig.   2, a  sonic  anemometer  turbulence  series and  its  IAAFT
surrogate  are  shown.   By  construction,  they  have  the  same  pdf
(bottom-right  plot  of   Fig.   2)  and  virtually  indistinguishable
autocorrelation  function  (bottom-left   plot  of  Fig.   2).   Basic
properties  of the original  turbulence series  and its  surrogate are
provided in Table  I. $T_i$ and $L_i$ denote  integral time and length
scales, respectively.
\begin{subequations}
\begin{eqnarray}
T_i = \int_0^{\infty} R(\tau) d\tau\\ L_i = U\cdot T_i
\end{eqnarray}
\end{subequations}
where  $R(\tau)$ is the  autocorrelation function.   From Fig.   2 and
Table I,  we can  safely infer that  the IAAFT surrogate  captures the
integral  pdf and  autocorrelation function  of the  original velocity
series  rather accurately.   Later on,  we  will show  that the  IAAFT
surrogates do not have the ability to capture the scale-dependent pdfs
of  velocity  increments  and  this forms the basis for the proposed 
intermittency hypothesis-testing.

\begin{widetext}
\begin{figure*}[ht]
\includegraphics[width=2.5in]{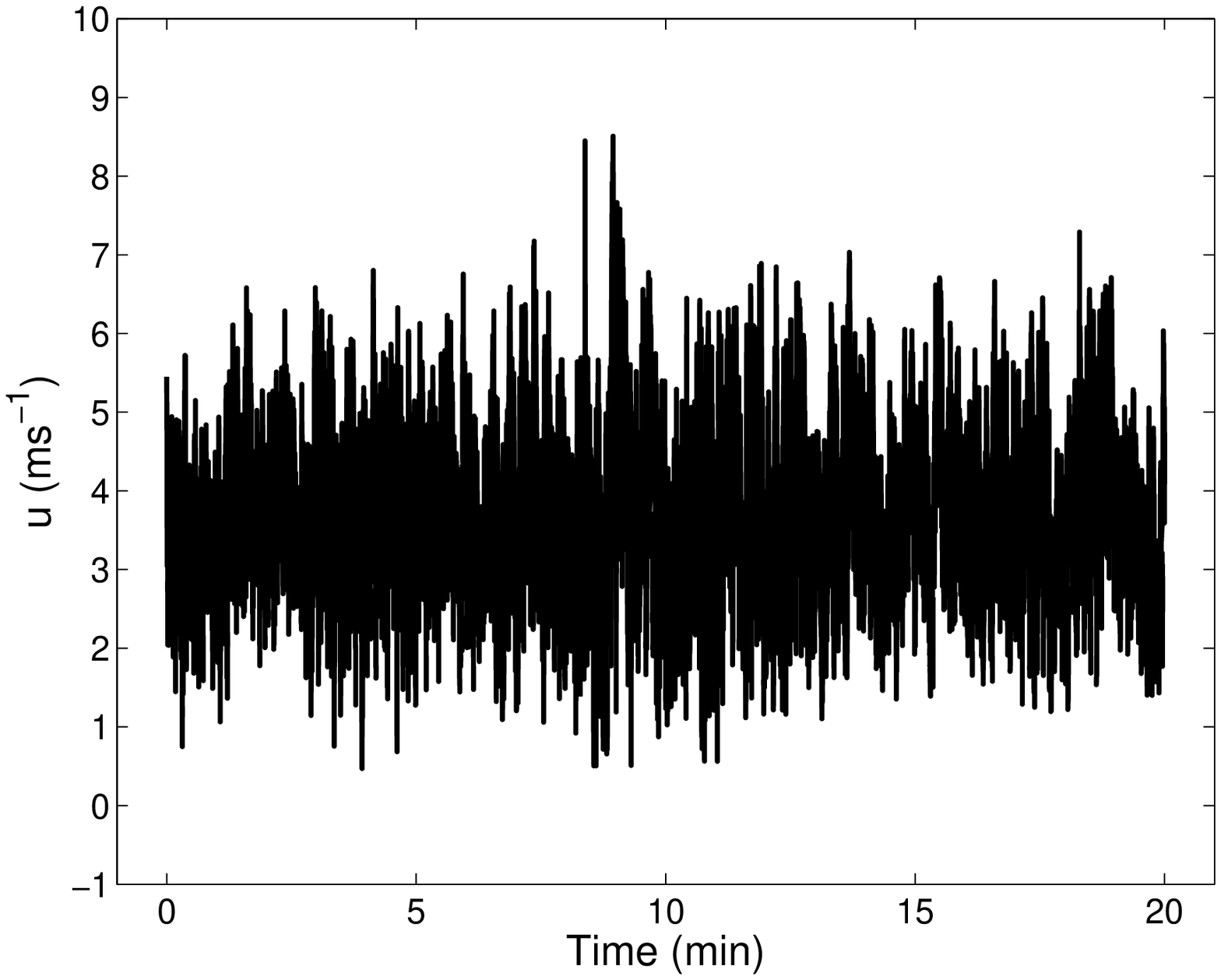}
\includegraphics[width=2.5in]{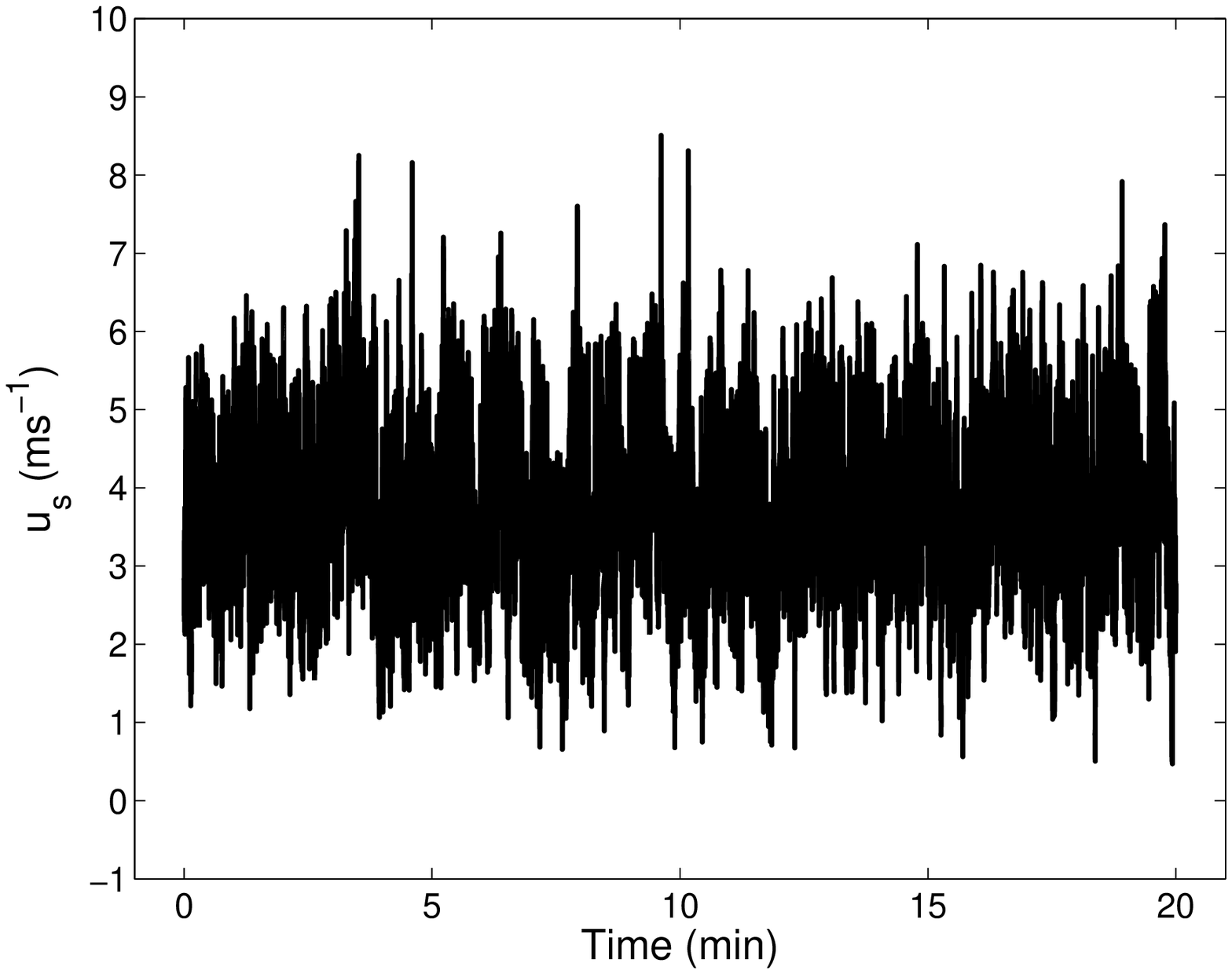}\\
\includegraphics[width=2.5in]{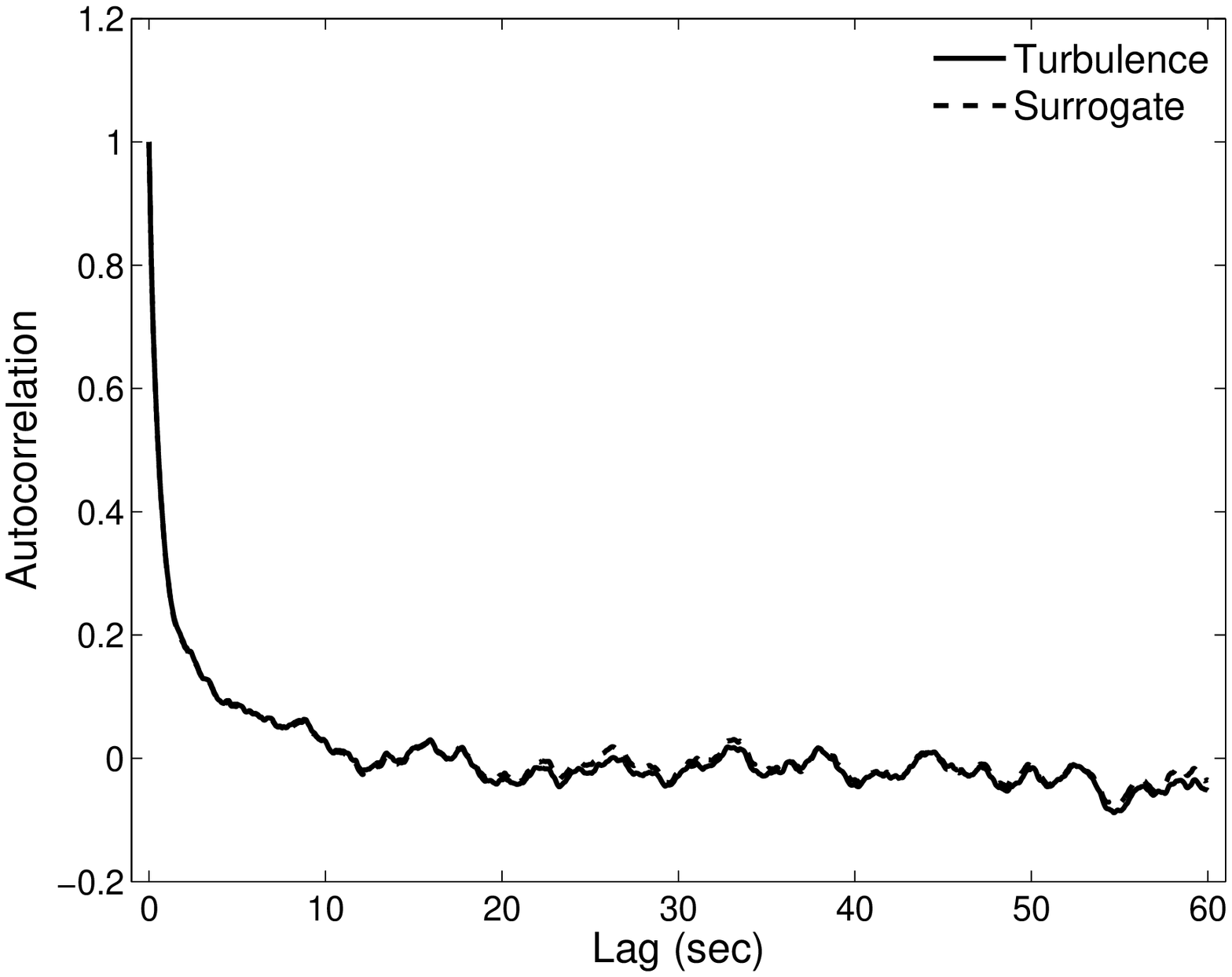}
\includegraphics[width=2.5in]{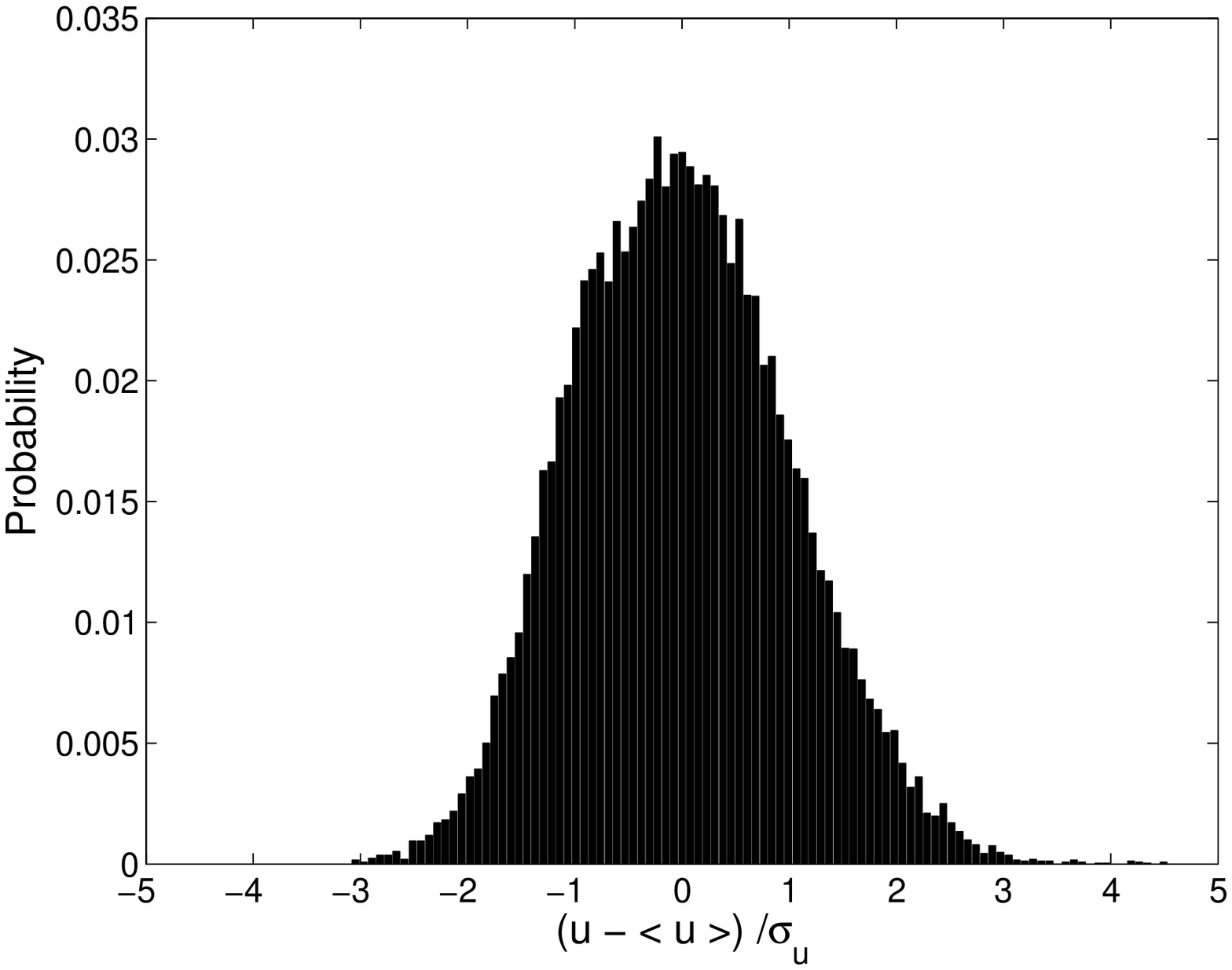}\\
\caption{Longitudinal  velocity  time   series  measured  by  a  sonic
anemometer  (top-left) and  its  surrogate series  generated by  IAAFT
methodology  (top-right).   Both series  have  approximately the  same
autocorrelation (bottom-left) and exactly the same probability density
function (bottom-right).}
\end{figure*}
\end{widetext}

\begin{widetext}
\begin{table*}[ht]
\caption{Basic Statistics of a  Sonic Anemometer Turbulence Series and
its Surrogate}\label{T1}
\begin{tabular}{cccccccc} \hline
Series Type  & $U$~(ms$^{-1}$) & $\sigma_u$~(ms$^{-1}$) &  $T_i$ (s) &
$L_i$  (m) \\  \hline  Turbulence  & 3.73  &  1.06 &  1.47  & 5.47  \\
Surrogate & 3.73 & 1.06 & 1.46 & 5.43 \\ \hline
\end{tabular} 
\end{table*}
\end{widetext}

\section{Description of Data}

In  this study,  we primarily  made  use of  an extensive  atmospheric
boundary  layer  turbulence dataset  (comprising  of sonic  anemometer
measurements)  collected  by various  researchers  from Johns  Hopkins
University, and the University  of California-Davis during Davis 1994,
1995, 1996,  1999 field  studies.  Comprehensive description  of these
field  experiments  (e.g.,   surface  cover,  fetch,  instrumentation,
sampling  frequency)  can  be  found  in  Pahlow  et  al.\cite{pahl01}
Briefly, the  collective attributes of  the field dataset  explored in
this study  are as follows: (i)  surface cover: bare  soil, and beans;
(ii) sampling frequency: 18 to 21  Hz; (iii) sampling period: 20 to 30
minutes; (iv) sensor height ($z$): 0.96 to 4.28 m.

The   ABL  field   measurements   are  seldom   free  from   mesoscale
disturbances, wave  activities, nonstationarities etc.   The situation
could be further  aggravated by several kinds of  sensor errors (e.g.,
random spikes, amplitude  resolution error, drop outs, discontinuities
etc.). Thus, stringent quality control and preprocessing of field data
is  of utmost importance  for any  rigorous statistical  analysis. Our
quality control  and preprocessing strategies are  described in detail
in   Basu  et   al.\cite{basu06}   After  the   quality  control   and
preprocessing  steps, we  were left  with 139  `reliable' near-neutral
($|z/L| \le 0.05$, where $z$ is  the sensor height and $L$ denotes the
Monin-Obukhov length)  sets of  runs for estimating  the intermittency
exponents.

We also  estimated $\mu$  from a fast-response  (10 kHz) hot  wire ABL
turbulence series  utilizing the magnitude cumulant  analysis. The hot
wire  measurements were  taken  at the  Surface  Layer Turbulence  and
Environmental Science  Test (SLTEST)  facility located in  the western
Utah   Great   Salt  Lake   desert   under  near-neutral   atmospheric
condition.\cite{maru03,kunk06} In the  following section, we will show
that the intermittency exponent  and other relevant statistics derived
from  this  high  Reynolds  number  ($Re$) hot  wire  measurement  are
surprisingly  similar  to  various  published  lower  $Re$  laboratory
experimental findings, and serve as benchmarks in the present study.

\begin{widetext}
\begin{table*}[ht]
\caption{Mean      Flow     Characteristics      of      the     Field
Measurements}\label{T2}
\begin{tabular}{cccccccc} \hline
Sensor  Type & $z$  (m) &  $U$~(ms$^{-1}$) &  $\sigma_u$~(ms$^{-1}$) &
$T_i$ (s) &  $L_i$ (m) \\ \hline  Hot wire Anemometer & 2.01  & 5.99 &
0.74 & 5.71 & 34.22 \\  Sonic Anemometer & $0.96-4.28$ & $1.60-7.30$ &
$0.34-1.57$ & $1.08-9.09$ & $2.57-33.22$ \\ \hline
\end{tabular} 
\end{table*}
\end{widetext}

Mean flow characteristics  of all the field measurements  are given in
Table II. For all the analyses, we have invoked Taylor's hypothesis to
convert time series to spatial series.

\section{Results}

\subsection{Analysis of Hot Wire Measurements}

In  this  section, hot  wire  measurements  and  their surrogates  are
analyzed  to: (a) demonstrate  the ability  of the  magnitude cumulant
analysis  to  accurately   estimate  the  intermittency  structure  of
turbulent  velocity  series,  and  (b) establish  that  the  surrogate
series, while  preserving the pdf  and spectrum of the  original data,
destroy the intermittency structure.

\begin{widetext}
\begin{figure*}[ht]
\includegraphics[width=2.5in]{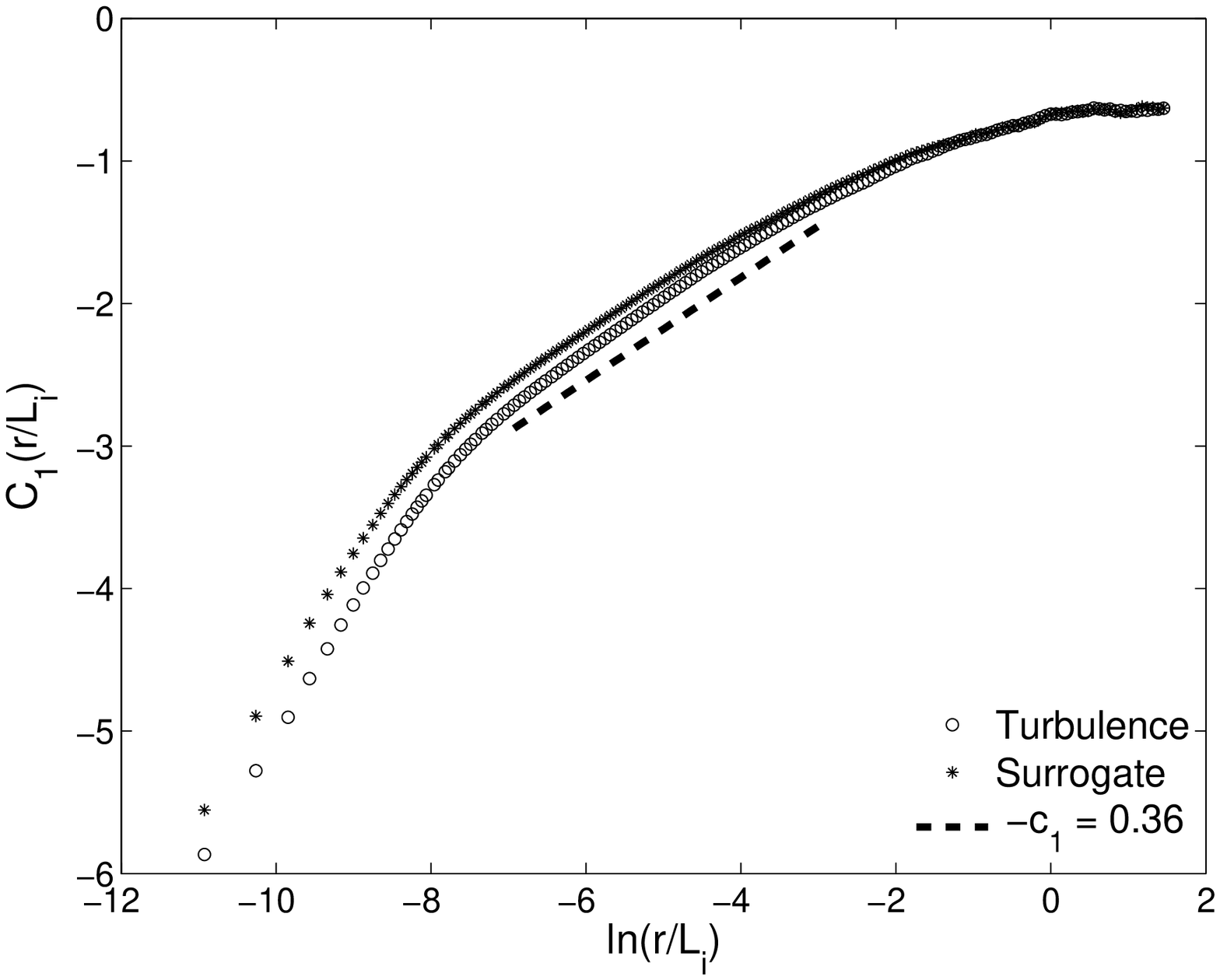}
\includegraphics[width=2.5in]{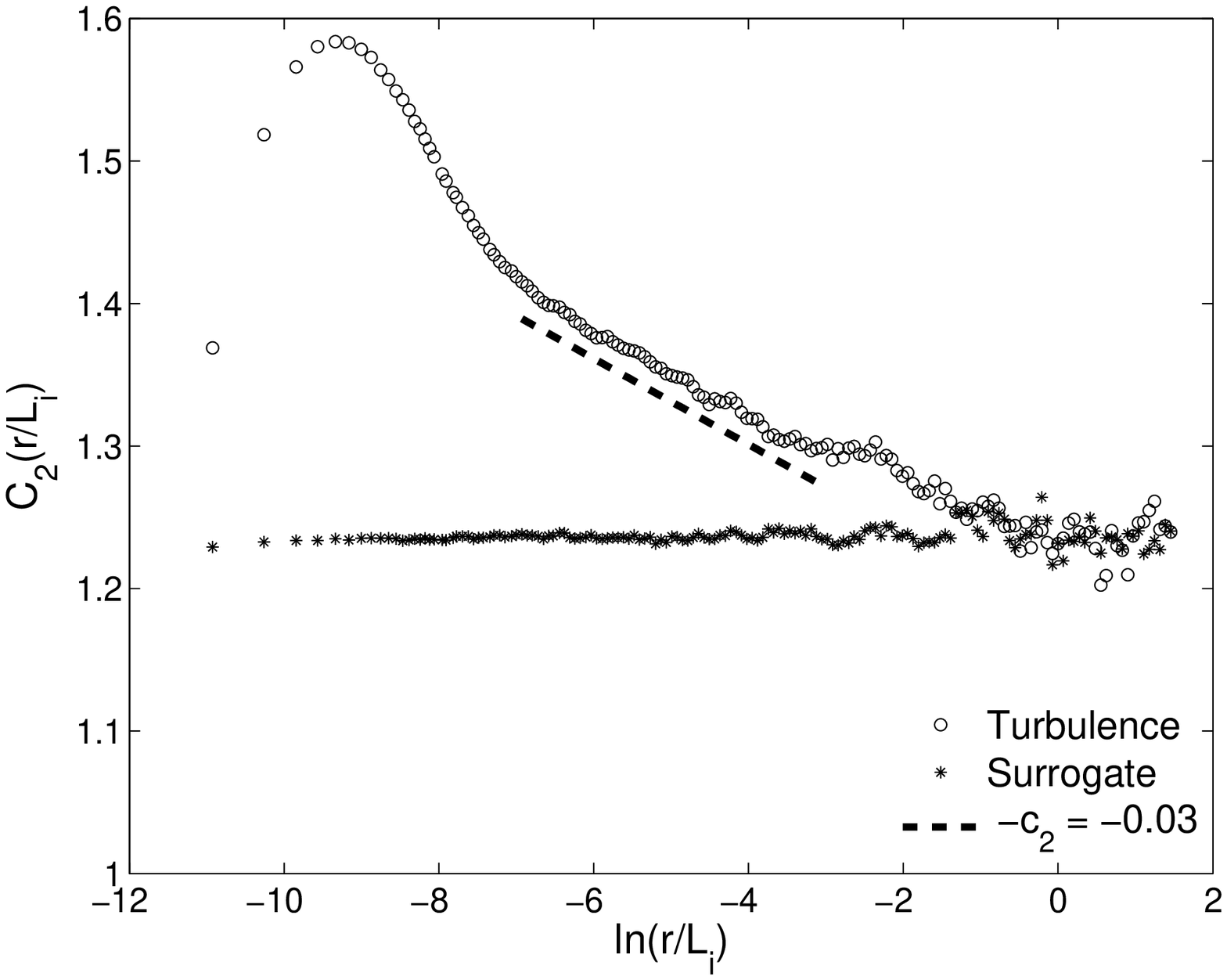}\\
\includegraphics[width=2.5in]{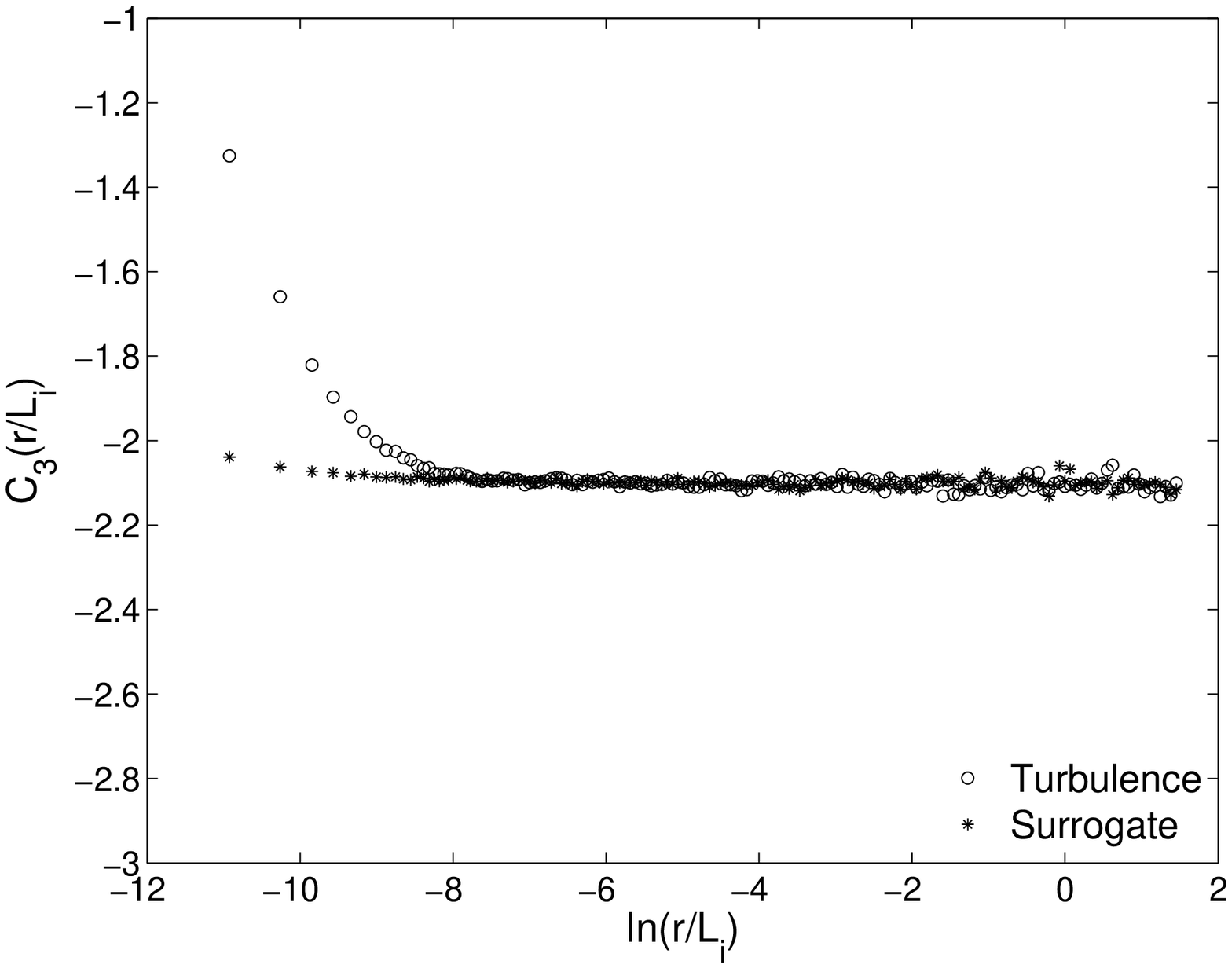}
\includegraphics[width=2.5in]{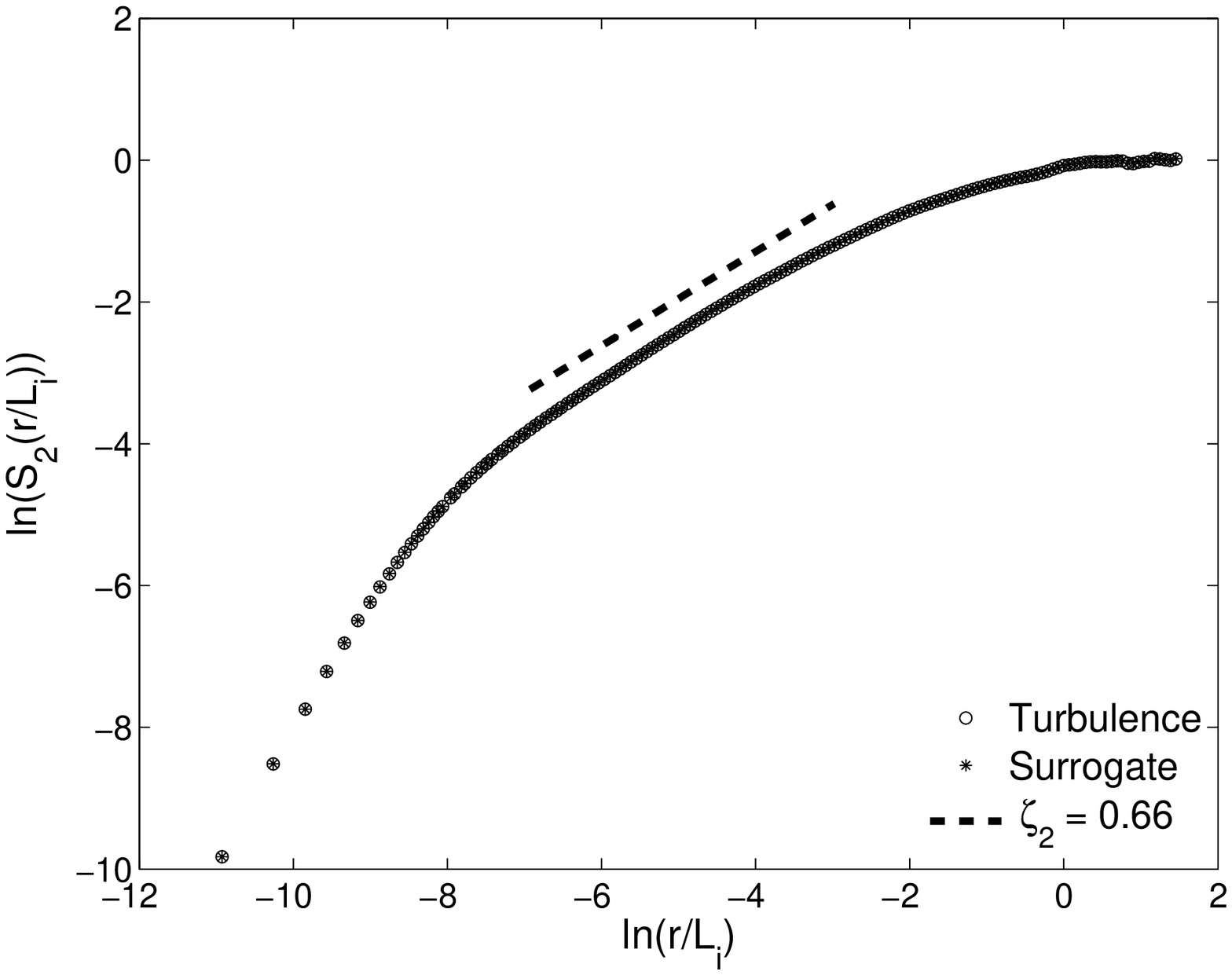}\\
\caption{$C_1(r/L_i)$   (top-left),   $C_2(r/L_i)$   (top-right)   and
$C_3(r/L_i)$ (bottom-left) computed using the hot wire measurements of
Kunkel and  Marusic.\cite{kunk06} The second-order  structure function
is  also shown  (bottom-right).  Extended  range of  inertial-range is
evident  in all  the  sub-plots.  The  circles  denote the  statistics
corresponding  to  the  original   turbulence  series  and  the  stars
represent  the statistics  computed from  the IAAFT  surrogate series.
Clearly, the original series  portray the signatures of a multifractal
process.   In  contrast,  the  surrogate  series shows  the  signs  of
monofractality.   The dashed  lines show  the slopes  $  -c_1^{turb} =
0.36$   (top-left),   $   -c_2^{turb}   =  -0.03$   (top-right),   and
$\zeta_2^{turb} = 0.66$ (bottom-right), respectively.}
\end{figure*}
\end{widetext}

In  Fig.   3, the  magnitude  cumulants  computed  from the  hot  wire
measurements  of  Kunkel  and  Marusic\cite{kunk06} are  shown.   This
turbulence series is  30 min long ($\sim 18$  million data points) and
captures  scales   down  to  the  Kolmogorov   scale.   The  following
observations can be made from Fig. 3:
\begin{itemize}
\item $c_1^{turb}$ computed from the turbulence velocity signal is
close to $-0.36$.  It agrees quite well with the existing results from
low $Re$ laboratory experiments.\cite{delo01,chev06}
 
\item $c_2^{turb}$  is approximately  equal to $0.03$.   Delour et
al.\cite{delo01}  and Chevillard et  al.\cite{chev06} reported  $c_2 =
0.025 \pm  0.003$ based  on several experiments  and claimed it  to be
`universal'.  From  Eq. 7, we  can compute the  intermittency exponent
$\mu  \simeq  9\cdot  c_2^{turb}  \simeq 0.27$.   In  the  literature,
researchers    have   reported    $\mu$   ranging    from    0.18   to
0.7.\cite{sree93,pras97} In  the case of atmospheric  data, the `best'
direct  estimate  is $0.25  \pm  0.05$\cite{sree93}  and our  indirect
magnitude cumulant analysis-based result is in agreement with it.

\item  From  Eq.  6,  we  can  derive:  $\zeta_2^{turb} =  -2\cdot
c_1^{turb}-2\cdot c_2^{turb} = 2\cdot 0.36 -2\cdot 0.03 = 0.66$.  Fig.
3 (bottom-right) shows the  second-order structure function. The slope
of this  plot gives  $\zeta_2^{turb} = 0.66$.   Thus, our  results are
self-consistent.

\item In the inertial range,  $c_3^{turb}$ seems to be zero.  This
indicates that the statistics  of the velocity increments are possibly
log-normal.\cite{delo01,chev06}

\item $c_2^{surr}$  estimated from  the surrogate of  the measured
turbulence  velocity series  is zero,  i.e., the  surrogate  series is
non-intermittent.

\item  By construction,  the surrogate series (i.e.,  turbulence without
intermittency)    preserves   the   second-order    statistics.    So,
$\zeta_2^{turb} =  \zeta_2^{surr}$. Using this  relationship, the fact
that $c_2^{surr} = 0$, and Eq.  6, it is straightforward to show that:
$c_1^{turb}  +   c_2^{turb}  =  c_1^{surr}$.  In   the  present  case this
rsults in:
$c_1^{surr}  =  -0.36 +  0.03  = -0.33$  (see  also  Fig.  3).   Thus,
$\zeta_1^{surr}  = -c_1^{surr}  = 0.33$  is  in full  accord with  K41
hypothesis of $\zeta_1 = 1/3$.\cite{fris95}

The   relationship  $c_1^{turb}   +  c_2^{turb}   =   c_1^{surr}$  has
significant practical implication.  It insinuates that one can roughly
estimate  $c_2^{turb}$  (and  thus  $\mu$)  by  means  of  first-order
magnitude  cumulants of  turbulence and  its  corresponding surrogate,
i.e.,  $c_2^{turb} = c_1^{turb}  - c_1^{surr}$.   In our  opinion, for
estimating  intermittency in  short length  geophysical  signals, this
simple  indirect  method  which  does not  require  even  second-order
magnitude cumulant computation would be quite useful.

\item In  the turbulence literature, there is  a general consensus
that  $\zeta_3 =  1$.  From  our results,  we  find $\zeta_3^{turb}  =
-3\cdot c_1^{turb}  - 9/2\cdot c_2^{turb}  = 1.08-0.135 =  0.95$, very
close to the well-accepted value.

\end{itemize}


We proceed further by comparing  the pdf of the velocity and surrogate
increments using the skewness,  asymmetry factor, and flatness defined
as:
\begin{subequations}
\begin{equation}
Skewness(r) =  \frac {\langle  (\Delta u)^3 \rangle}  {\langle (\Delta
u)^2 \rangle^{3/2}}
\end{equation}
\begin{equation}
Asymmetry(r) = \frac {\langle  (\Delta u)^3 \rangle} {\langle |(\Delta
u)^3| \rangle}
\end{equation}
\begin{equation}
Flatness(r) =  \frac {\langle  (\Delta u)^4 \rangle}  {\langle (\Delta
u)^2 \rangle^{2}}
\end{equation}
\end{subequations}

From  Fig.  4  (left),  it  is evident  that  the original  turbulence
increment series show  negative skewness (up to $\sim  0.6$) for small
scales  in accord with  the existing  literature (e.g.,  Chevillard et
al.\cite{chev06}). This negative skewness is believed to be related to
the   vortex   folding  and   stretching   process.   Mal\'{e}cot   et
al.\cite{male00}  argued that the  asymmetry factor  (see Eq.   10 for
definition) is a  better measure of the asymmetry of  the pdf than the
skewness. We found that both  of these signed odd-order moments behave
quite similarly  (Fig. 4, left).  The origin  of spurious oscillations
of these odd-order moments for  large scales ($ln(r/L_i) > -4$) is not
well understood.   The flatness plot  (Figure 4, right)  also portrays
anticipated characteristics.   Flatness corresponding to  the integral
scale is  close to  3 (hallmark of  Gaussian velocity  increments) and
becomes  exceedingly  large  for  smaller scales.   In  contrast,  the
surrogate  shows  Gaussian   characteristics  for  all  scales.   This
corroborates  the fact  that  surrogates cannot  capture  the pdfs  of
turbulence velocity increments.

\begin{widetext}
\begin{figure*}[ht]
\includegraphics[width=2.5in]{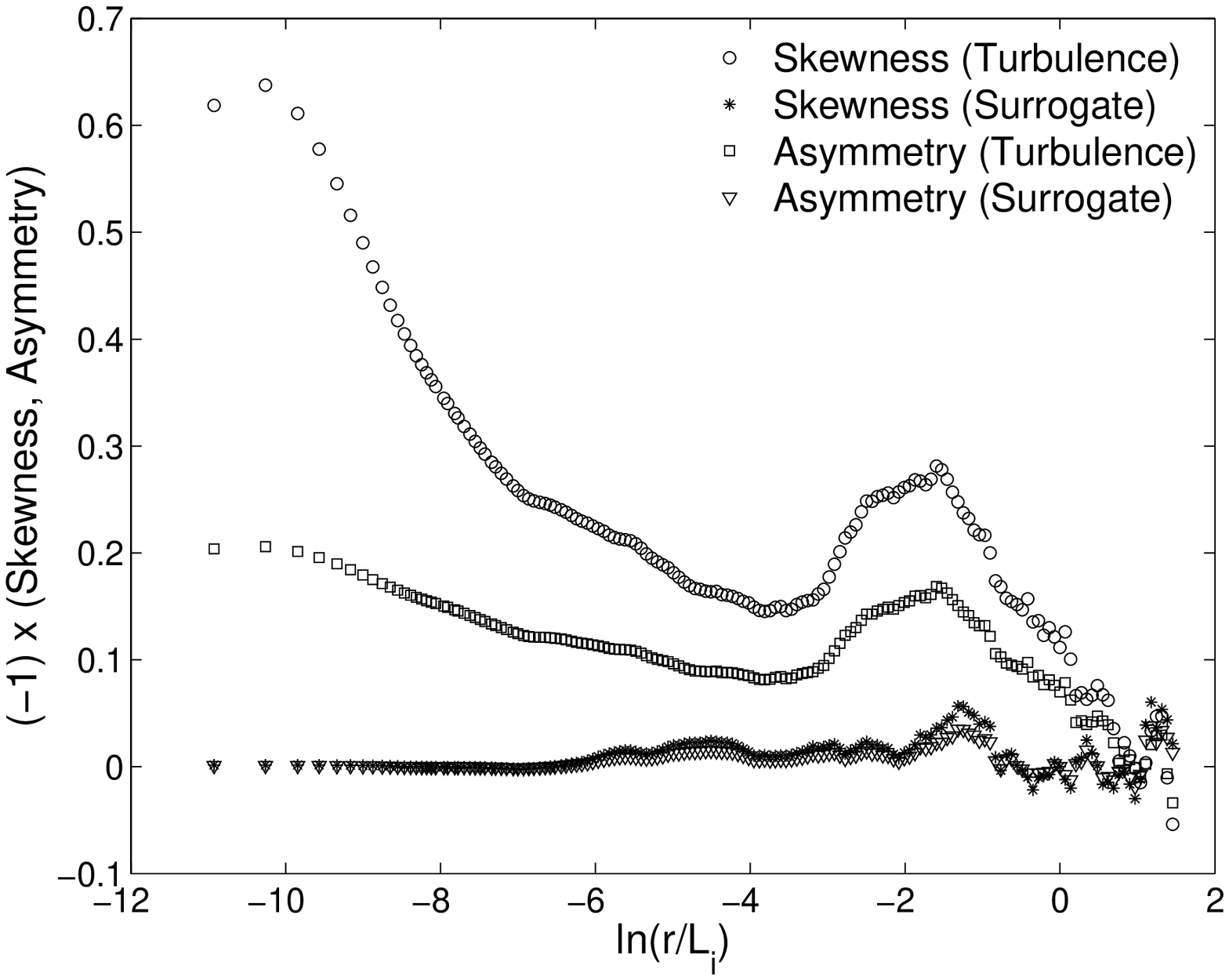}
\includegraphics[width=2.5in]{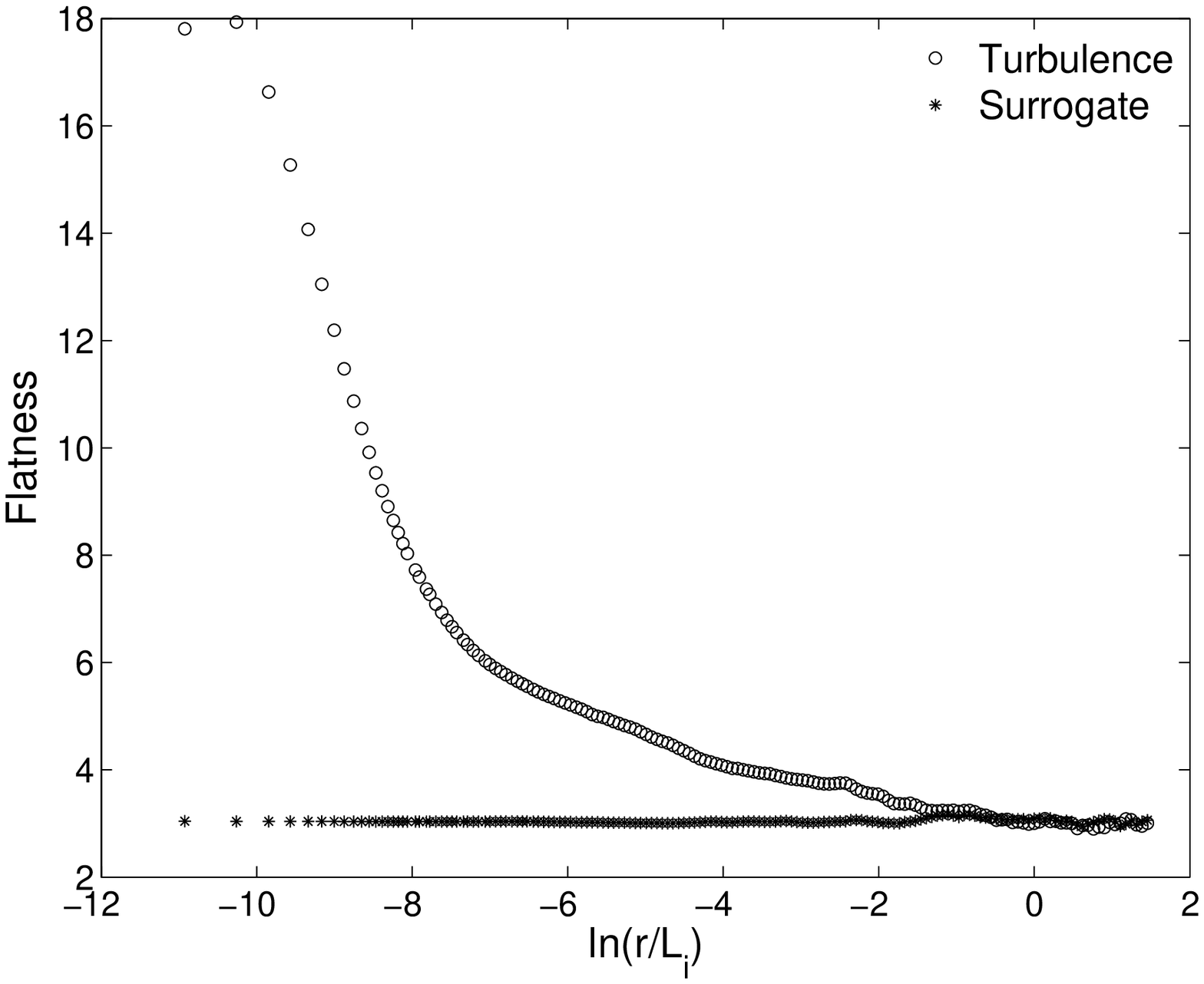}
\caption{Negative  skewness,  asymmetry  factor (left),  and  flatness
(right) of the longitudinal  velocity increments and the increments of
the surrogate series.  We utilized the hot wire measurements of Kunkel
and Marusic.\cite{kunk06}}
\end{figure*}
\end{widetext}

\subsection{Analysis of Sonic Anemometer Measurements}

Magnitude cumulants and second-order structure functions computed from
a sonic  anemometer series are shown  in Fig.  5. The  trends are very
similar to Fig. 3, albeit quite noisy. From this figure, we calculated
$c_1 = -0.37$,  $c_2 = 0.06$, and $\zeta_2 = 0.63$.   We would like to
emphasize  that  even in  this  short  time  series scenario,  we  can
reliably detect  intermittency with the  help of IAAFT  surrogate (see
Fig. 5 top-right). Admittedly, the estimation of $c_2$ is possibly not
very accurate.   The estimation can  be improved by using  quenched or
annealed averaging strategy (discussed below).

\begin{widetext}
\begin{figure*}[ht]
\includegraphics[width=2.5in]{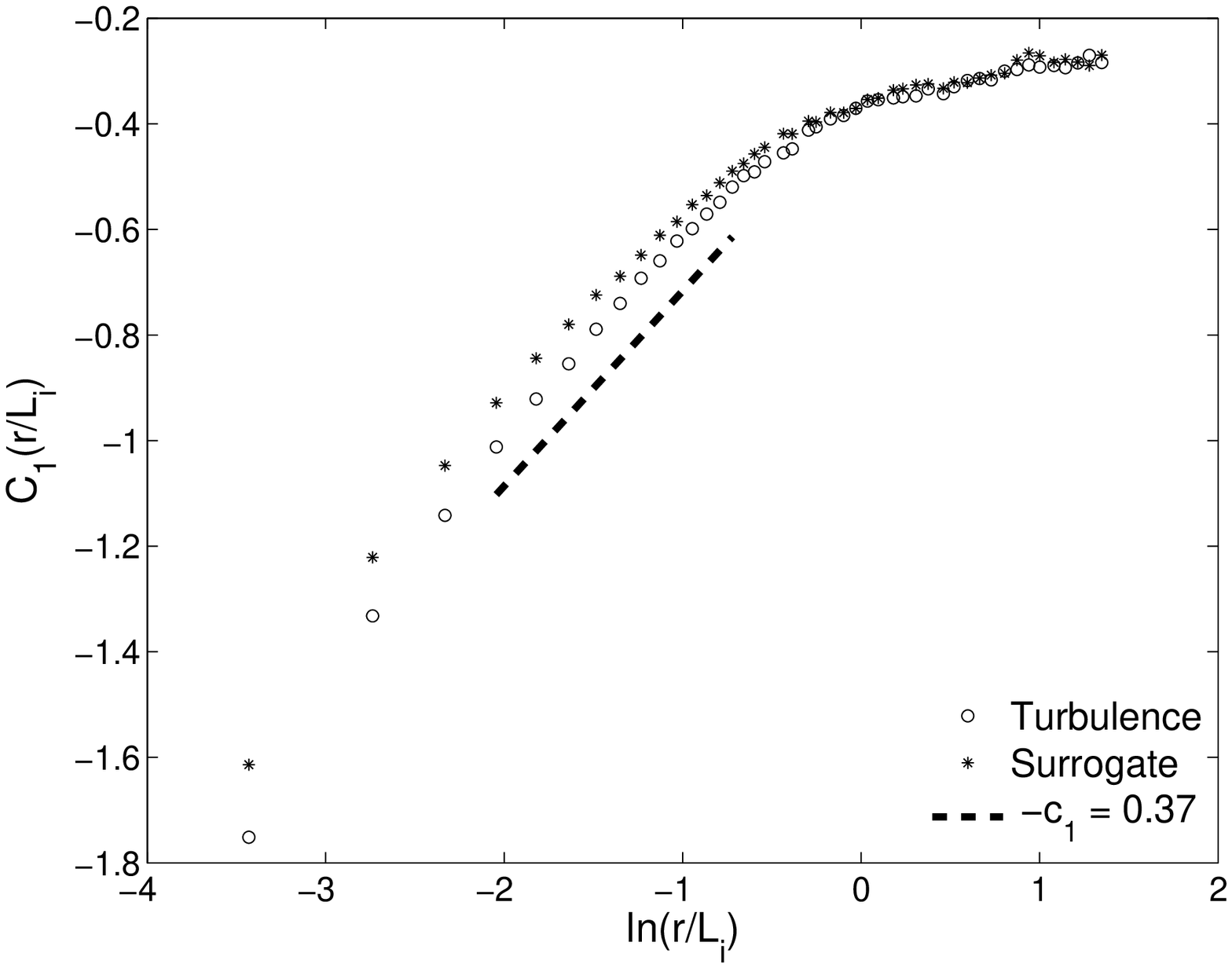}
\includegraphics[width=2.5in]{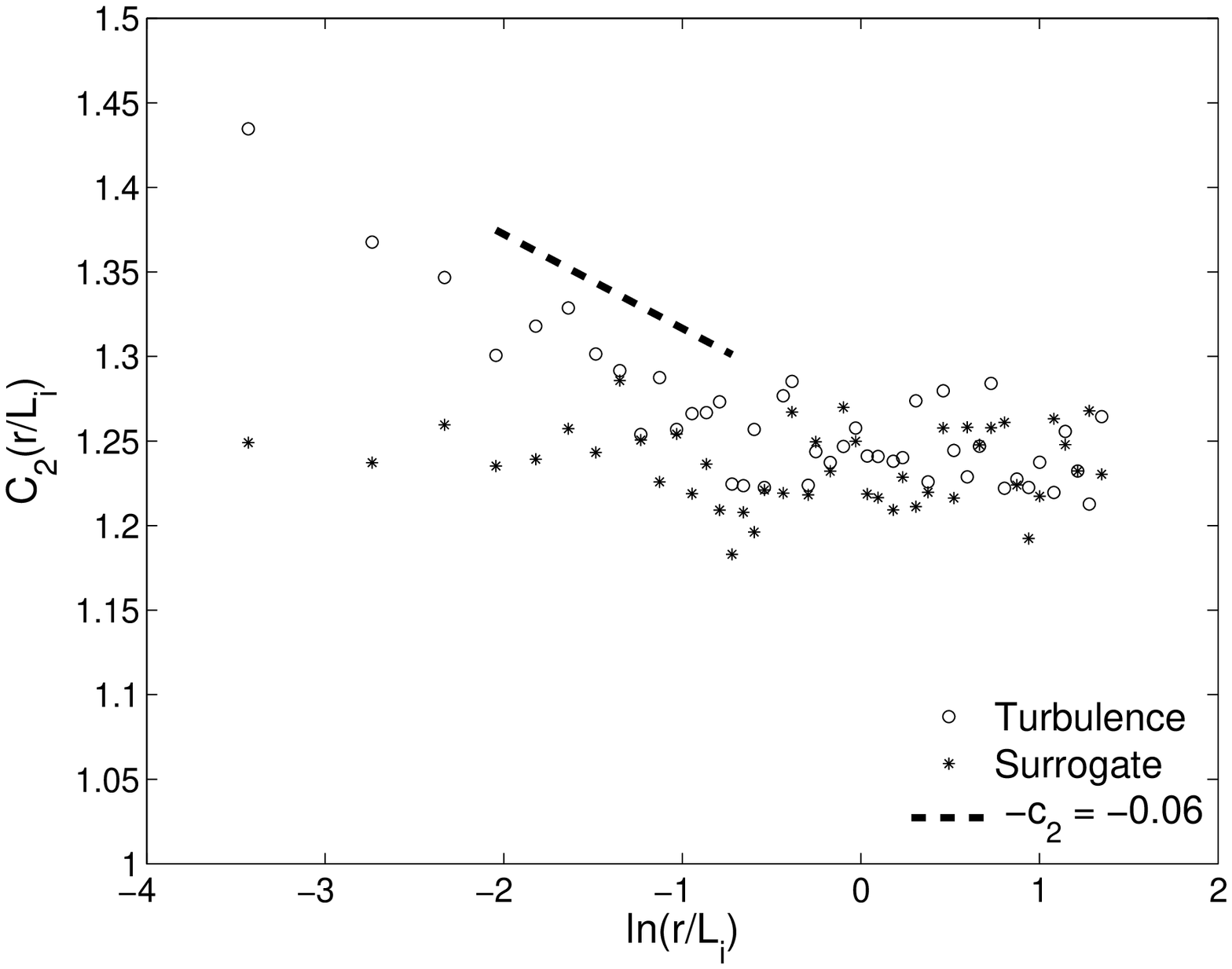}\\
\includegraphics[width=2.5in]{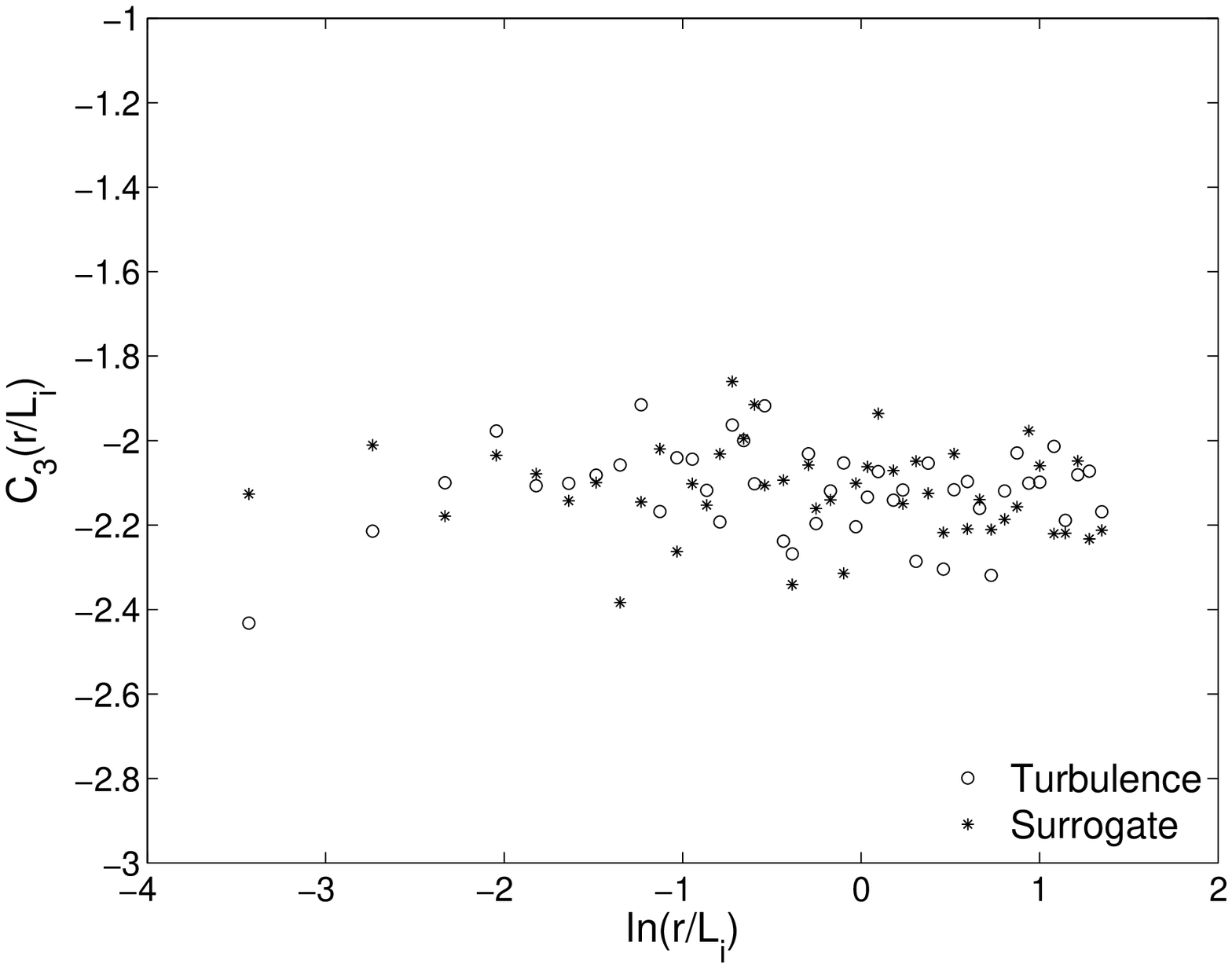}
\includegraphics[width=2.5in]{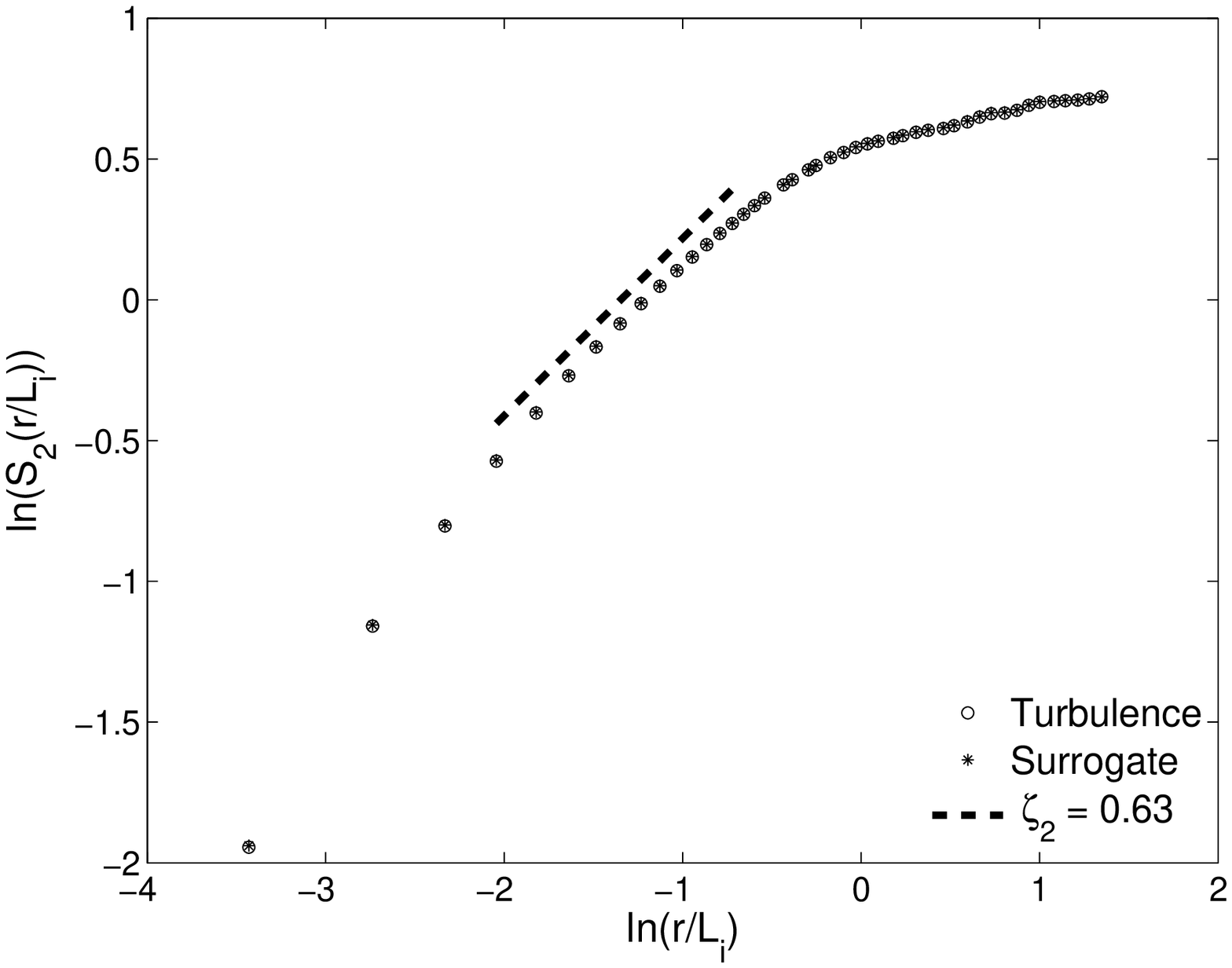}\\
\caption{Same  as Figure  3,  but estimated  from  a sonic  anemometer
series.}
\end{figure*}
\end{widetext}


It  is quite  difficult  to manually  yet  objectively select  scaling
ranges from a large  dataset (specifically 139 near-neutral turbulence
series).   So,  we  used   an  automated  scaling  range  of  $[4\cdot
U/f_s~L_i/2]$  for  individual series.   Here  $U$,  $f_s$, and  $L_i$
denote mean  velocity, sampling frequency, and  integral length scale,
respectively.   From  each  individual  turbulence  series  and  their
surrogates, we  calculated the  corresponding $c_1$ and  $c_2$ values.
Subsequently, from these 139 $c_1^{turb}$, $c_2^{turb}$, $c_1^{surr}$,
and  $c_2^{surr}$   combinations,  we  computed   the  best  estimates
(ensemble  mean  plus/minus   one  standard  deviation)  as:  $\langle
c_1^{turb}  \rangle  = -0.33\mp0.03$,  $\langle  c_2^{turb} \rangle  =
0.038\pm  0.017$, $\langle  c_1^{surr} \rangle  = -0.30\mp  0.03$, and
$\langle  c_2^{surr}  \rangle   =  0.002\pm  0.013$.   This  averaging
strategy  is  similar  to   the  quenched  averaging  method  used  by
Arn\'{e}odo  et al.\cite{arne00} The  key result:  $\langle c_2^{turb}
\rangle >> \langle c_2^{surr}  \rangle$, without any doubt, once again
guarantees  that  the IAAFT  surrogates  can  be  faithfully used  for
turbulence intermittency detection testing.
	
As   an   alternative   strategy,   using   the   annealed   averaging
method\cite{arne00},  we  have  also   computed  the  average  of  the
magnitude  cumulants  (i.e.,  $\langle  C_1\left(r/L_i\right)\rangle$,
$\langle C_2\left(r/L_i\right)\rangle$ )  from the same turbulence and
surrogate  datasets (Fig.   6).   From this  figure,  we estimate  the
slopes as:  $\overline{c}_1^{turb}= -0.35$, $  \overline{c}_2^{turb} =
0.042$,  $ \overline{c}_1^{surr} =  -0.31$, $  \overline{c}_2^{surr} =
0.002$.  Obviously,  there is  no significant discrepancy  between the
quenched  and  annealed averaged  statistics,  as  would be  wishfully
expected.   Lastly, these statistics  highlight that  the relationship
established  in   Section  V-A,  i.e.,  $c_1^{turb}   +  c_2^{turb}  =
c_1^{surr}$, is also valid under annealed averaging.

\begin{widetext}
\begin{figure*}[ht]
\includegraphics[width=2.5in]{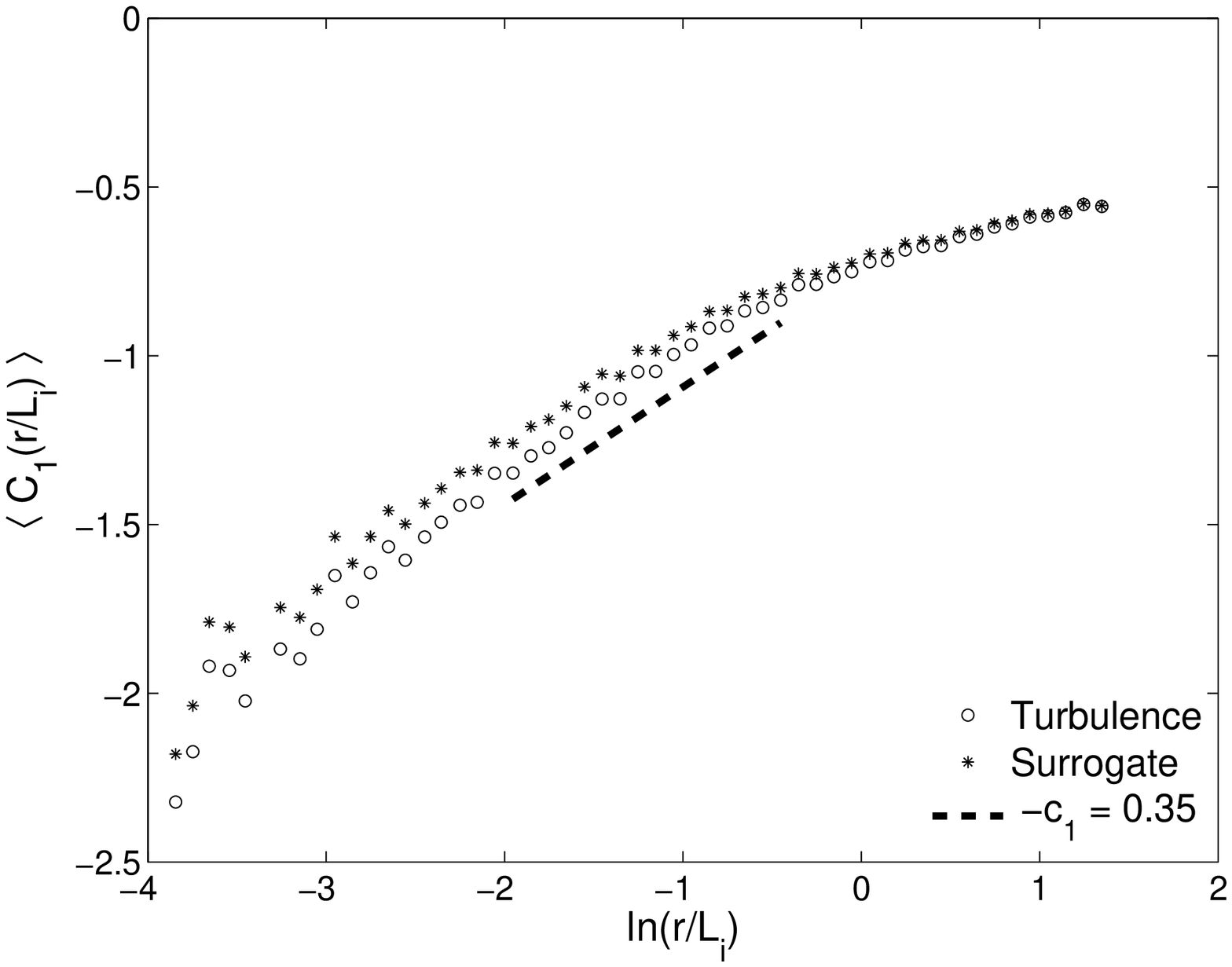}
\includegraphics[width=2.5in]{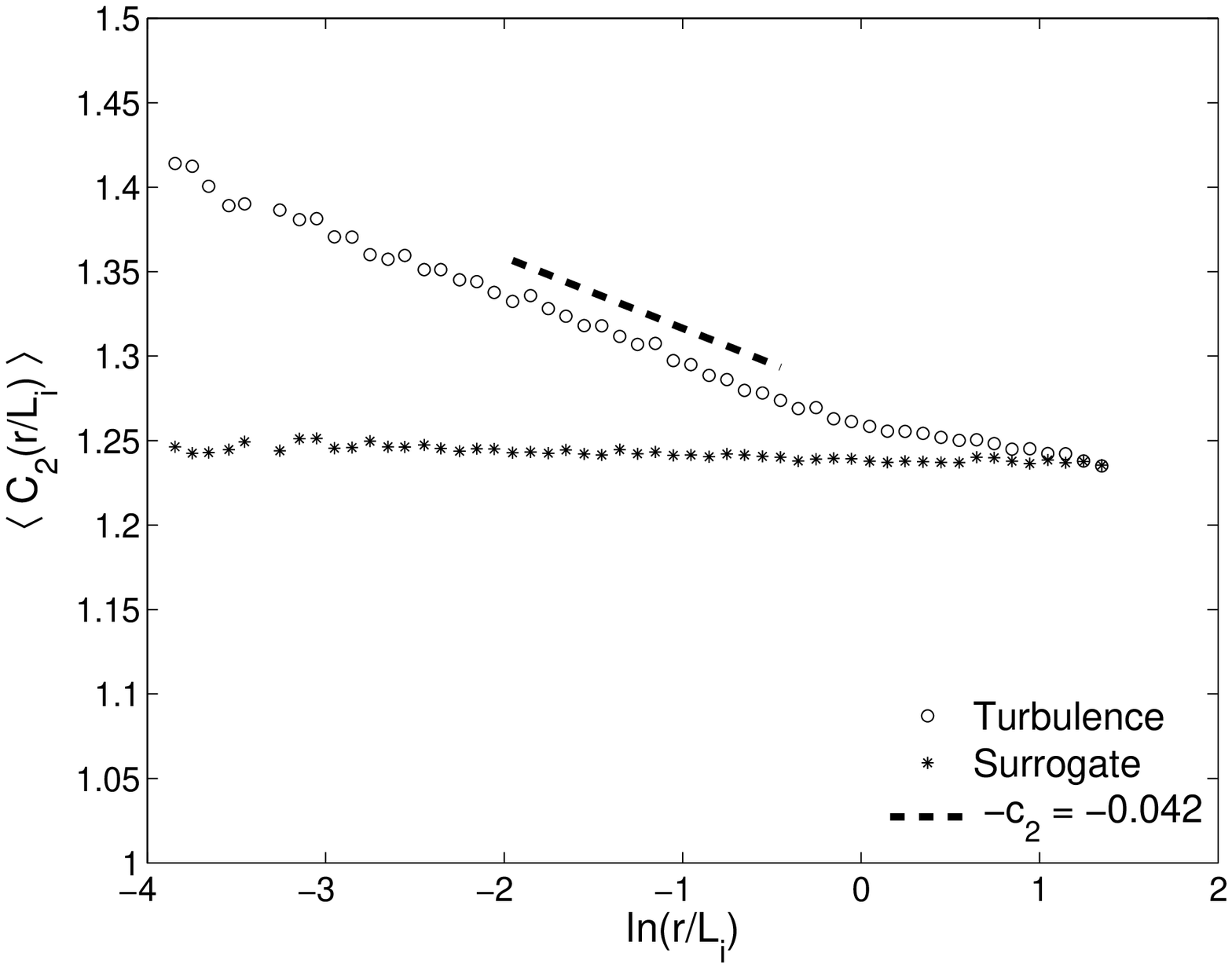}
\caption{Ensemble  averaged $C_1(r/L_i)$  and $C_2(r/L_i)$  plots from
139  near-neutral   turbulence  series  (circles)   and  corresponding
surrogates    (stars).    The   dashed    lines   show    the   slopes
$-\overline{c}_1^{turb} = 0.35$  (left), and $-\overline{c}_2^{turb} =
-0.042$ (right), respectively.}
\end{figure*}
\end{widetext}


\section{Concluding Remarks and Future Perspectives}

In  this work,  we have  established  a framework  based on  magnitude
cumulant and  surrogate analyses to  reliably detect and  estimate the
intermittency coefficient from short  turbulent time series. By virtue
of  this framework, ensemble  scaling results  extracted from  a large
number  of neutrally  stratified atmospheric  surface  layer turbulent
series  (predominantly acquired  by slow  response  sonic anemometers)
from  various  field   campaigns  remarkably  concur  with  well-known
published (mostly laboratory experimental) results.


The focus of the present study was on neutrally stratified atmospheric
turbulence.   However,  it  is  widely known  that  neutral  stability
conditions are rarely encountered in  the real atmosphere. Most of the
time,  the  atmospheric  boundary   layer  is  strongly  modulated  by
buoyancy.   It is  commonly  assumed that  the  effect of  atmospheric
stability  is felt  only at  the  `buoyancy range',  which has  scales
considerably  larger than  the inertial  range.  Recently,  Aivalis et
al.\cite{aiva02} studied the  intermittency behavior of temperature in
the convective  surface layer using cold wire  anemometry.  They found
that the classical inertial range remains intact in convective surface
layer  and the scaling  exponents approach  values appropriate  to the
intermittent case of isotropic  turbulence. They also noticed that the
scaling  exponents  corresponding to  the  buoyancy  range are  highly
anomalous.  In contrast, Shi et al.\cite{shi05} found that in the case
of temperature, the inertial range scaling exponents are unambiguously
impacted  by  atmospheric   stability.   However,  their  results  are
inconclusive in the case of velocity signals under different stability
regimes. We believe that the  unresolved question as to whether or not
the   inertial-range  intermittency   is  influenced   by  large-scale
anisotropic forcing  of atmospheric stability is  of great consequence
and needs  further consideration. Abundant  high-quality slow response
data  collected under  different atmospheric  regimes in  recent years
(e.g.,    Cooperative     Atmosphere-Surface    Exchange    Study    -
CASES99\cite{poul02})  could  be   coupled  with  the  robust  scaling
analysis and  estimation framework explored  in this study in order  to shed
new light into this fundamental problem.
	
\begin{acknowledgments}
We would like  to thank Gary Kunkel and Ivan  Marusic for providing us
with the hot wire measurements  taken at the SLTEST facility, Utah. We
are also  grateful to all the researchers  who painstakingly collected
data during the Davis field campaigns.  This work was partially funded
by  the  National  Science  Foundation  (ANT-0538453)  and  the  Texas
Advanced  Research   Program  (003644-0003-2006)  grants   awarded  to
S. Basu.  E. Foufoula-Georgiou acknowledges support by NASA, under its
Global  Precipitation  Mission  (GPM)  program,  and by  NSF  via  the
National  Center  for Earth-surface  Dynamics  (NCED) under  agreement
EAR-0120914.
\end{acknowledgments}


\begin{thebibliography}{0}
\expandafter\ifx\csname natexlab\endcsname\relax\def\natexlab#1{#1}\fi
\expandafter\ifx\csname bibnamefont\endcsname\relax
  \def\bibnamefont#1{#1}\fi
\expandafter\ifx\csname bibfnamefont\endcsname\relax
  \def\bibfnamefont#1{#1}\fi
\expandafter\ifx\csname citenamefont\endcsname\relax
  \def\citenamefont#1{#1}\fi
\expandafter\ifx\csname url\endcsname\relax
  \def\url#1{\texttt{#1}}\fi
\expandafter\ifx\csname urlprefix\endcsname\relax\def\urlprefix{URL }\fi
\providecommand{\bibinfo}[2]{#2}
\providecommand{\eprint}[2][]{\url{#2}}

\end{thebibliography}


\begin{thebibliography}{}
\expandafter\ifx\csname natexlab\endcsname\relax\def\natexlab#1{#1}\fi
\expandafter\ifx\csname                     bibnamefont\endcsname\relax
\def\bibnamefont#1{#1}\fi                       \expandafter\ifx\csname
bibfnamefont\endcsname\relax                 \def\bibfnamefont#1{#1}\fi
\expandafter\ifx\csname                    citenamefont\endcsname\relax
\def\citenamefont#1{#1}\fi \expandafter\ifx\csname url\endcsname\relax
\def\url#1{\texttt{#1}}\fi                      \expandafter\ifx\csname
urlprefix\endcsname\relax\def\urlprefix{URL                        }\fi
\providecommand{\bibinfo}[2]{#2}
\providecommand{\eprint}[2][]{\url{#2}}

\bibitem{fris95}  U.~Frisch,  {\it  Turbulence} (Cambridge  University
Press, Cambridge, UK, 1995), p. 296.

\bibitem{sche99}   D.    Schertzer,    S.    Lovejoy,   F.    Schmitt,
  Y. Chigirinskaya, D.  Marsan, \bibinfo{title} ``Multifractal cascade
  dynamics and  turbulent intermittency'', \bibinfo{journal}{Fractals}
  \textbf{\bibinfo{volume}{5}},                    \bibinfo{pages}{427}
  (\bibinfo{year}{1997}).

\bibitem{scot99}   A.~Scotti  and  C.~Meneveau,   \bibinfo{title}  ``A
  fractal  model  for  large  eddy  simulation  of  turbulent  flow'',
  \bibinfo{journal}{Physica     D}     \textbf{\bibinfo{volume}{127}},
  \bibinfo{pages}{198} (\bibinfo{year}{1999}).
  
\bibitem{velh01} H.   F.  Campos  Velho, R. R.   Rosa, F.   M.  Ramos,
  R. A. Pielke,  G. A. Degrazia, C. Rodrigues  Neto, and A. Zanandrea,
  \bibinfo{title}  ``Multifractal  model   for  eddy  diffusivity  and
  counter-gradient      term     in      atmospheric     turbulence'',
  \bibinfo{journal}{Physica     A}     \textbf{\bibinfo{volume}{295}},
  \bibinfo{pages}{219} (\bibinfo{year}{2001}).
 
\bibitem{basu04} S.~Basu, E.~Foufoula-Georgiou, and F.~Port\'{e}-Agel,
  \bibinfo{title}  ``Synthetic turbulence, fractal  interpolation, and
  large-eddy    simulation'',   \bibinfo{journal}{Phys.     Rev.    E}
  \textbf{\bibinfo{volume}{70}},                \bibinfo{pages}{026310}
  (\bibinfo{year}{2004}).
  
\bibitem{anto05}  M.~Antonelli,  M. Martins  Afonso,  A. Mazzino,  and
  U.~Rizza, \bibinfo{title} ``Structure of temperature fluctuations in
  turbulent   convective    boundar   layers'',   \bibinfo{journal}{J.
  Turbulence}    \textbf{\bibinfo{volume}{6}},    \bibinfo{pages}{DOI:
  10.1080/14685240500332049} (\bibinfo{year}{2005}).
  
\bibitem{burt05} G.  C.  Burton and  W.  J.  A.  Dahm, \bibinfo{title}
  ``Multifractal  subgrid-scale  modeling  for large-eddy  simulation.
  {I}.      Model    development     and    a     priori    testing'',
  \bibinfo{journal}{Phys.     Fluids}   \textbf{\bibinfo{volume}{17}},
  \bibinfo{pages}{075111} (\bibinfo{year}{2005}).

\bibitem{moni75}  A.~Monin  and   A.~Yaglom,  {\it  Statistical  Fluid
Mechanics} (MIT Press, Cambridge, MA, 1975), Vol. 2, p. 874.

\bibitem{mene91} C. Meneveau and K. R. Sreenivasan, \bibinfo{title}
  ``The multifractal nature of the turbulent energy dissipation'',
  \bibinfo{journal}{J. Fluid Mech.}    \textbf{\bibinfo{volume}{224}},
  \bibinfo{pages}{429} (\bibinfo{year}{1991}).

\bibitem{sree93} K. R.  Sreenivasan and P. Kailasnath, \bibinfo{title}
  ``An   update  on  the   intermittency  exponent   in  turbulence'',
  \bibinfo{journal}{Phys.     Fluids}    \textbf{\bibinfo{volume}{5}},
  \bibinfo{pages}{512} (\bibinfo{year}{1993}).

\bibitem{pras97}  A.   Praskovsky   and  S.   Oncley,  \bibinfo{title}
  ``Comprehensive measurements  of the intermittency  exponent in high
  Reynolds  number  turbulent  flows'', \bibinfo{journal}{Fluids  Dyn.
  Res.}       \textbf{\bibinfo{volume}{21}},      \bibinfo{pages}{331}
  (\bibinfo{year}{1997}).

\bibitem{clev04} J.   Cleve, M.  Greiner,  B.  R.  Pearson, and  K. R.
  Sreenivasan,   \bibinfo{title}  ``Intermittency   exponent   of  the
  turbulent   energy  cascade'',  \bibinfo{journal}{Phys.    Rev.   E}
  \textbf{\bibinfo{volume}{69}},                \bibinfo{pages}{066316}
  (\bibinfo{year}{2004}).

\bibitem{anse84} F.   Anselmet, Y.  Gagne,  E.  J.  Hopfinger,  and R.
  A.    Antonia,  \bibinfo{title}   ``High-order   velocity  structure
  functions  in turbulent  shear flows'',  \bibinfo{journal}{J.  Fluid
  Mech.}       \textbf{\bibinfo{volume}{140}},     \bibinfo{pages}{63}
  (\bibinfo{year}{1984}).
  
\bibitem{cham84} A.  J.  Chambers and R.  A.  Antonia, \bibinfo{title}
  ``Atmospheric      estimates      of     power-law      exponents'',
  \bibinfo{journal}{Boundary-Layer                           Meteorol.}
  \textbf{\bibinfo{volume}{28}},                   \bibinfo{pages}{343}
  (\bibinfo{year}{1984}).
  
\bibitem{maru03}  I.   Marusic  and  G.  J.   Kunkel,  \bibinfo{title}
  ``Streamwise   turbulence  intensity   formulation   for  flat-plate
  boundary       layers'',       \bibinfo{journal}{Phys.       Fluids}
  \textbf{\bibinfo{volume}{15}},                  \bibinfo{pages}{2461}
  (\bibinfo{year}{2003}).

\bibitem{kunk06} G. J. Kunkel  and I. Marusic, \bibinfo{title} ``Study
  of  the  near-wall-turbulent   region  of  the  high-Reynolds-number
  boundary  layer using  an  atmospheric flow'',  \bibinfo{journal}{J.
  Fluid  Mech.}   \textbf{\bibinfo{volume}{548}}, \bibinfo{pages}{375}
  (\bibinfo{year}{2006}).

\bibitem{fris78}  U.   Frisch,  P.-L.   Sulem,  and  M.   J.   Nelkin,
  \bibinfo{title}  ``A simple  dynamical model  of  intermittent fully
  developed    turbulence'',   \bibinfo{journal}{J.     Fluid   Mech.}
  \textbf{\bibinfo{volume}{87}},                   \bibinfo{pages}{719}
  (\bibinfo{year}{1978}).

\bibitem{delo01}  J.   Delour,  J.   F.  Muzy,  and  A.   Arn\'{e}odo,
  \bibinfo{title} ``Intermittency of {1D} velocity spatial profiles in
  turbulence:       {A}      magnitude       cumulant      analysis'',
  \bibinfo{journal}{Eur.  Phys.   J. B} \textbf{\bibinfo{volume}{23}},
  \bibinfo{pages}{243} (\bibinfo{year}{2001}).

\bibitem{schr96}  T.   Schreiber   and  A.   Schmitz,  \bibinfo{title}
  ``Improved     surrogate    data    for     nonlinearity    tests'',
  \bibinfo{journal}{Phys.  Rev.  Lett.} \textbf{\bibinfo{volume}{77}},
  \bibinfo{pages}{635} (\bibinfo{year}{1996}).

\bibitem{bohr98} T.~Bohr,  M. H. Jensen, G. Paladin,  and A. Vulpiani,
{\it Dynamical  Systems Approach to  Turbulence} (Cambridge University
Press, Cambridge, UK, 1998), p. 350.

\bibitem{lash04}   B.   Lashermes,   P.   Abry,   and   P.   Chainais,
  \bibinfo{title}  ``New  insights  into  the  estimation  of  scaling
  exponents'',  \bibinfo{journal}{International  Journal of  Wavelets,
  Multiresolution          and         Information         Processing}
  \textbf{\bibinfo{volume}{2}},                    \bibinfo{pages}{497}
  (\bibinfo{year}{2004}).

\bibitem{chev06} L.  Chevillard,  B. Castaing, E. L\'{e}v\^{e}que, and
  A.  Arn\'{e}odo,  \bibinfo{title} ``Unified multifractal description
  of velocity increments statistics in turbulence: {I}ntermittency and
  skewness'',               \bibinfo{journal}{Physica               D}
  \textbf{\bibinfo{volume}{218}},                   \bibinfo{pages}{77}
  (\bibinfo{year}{2006}).

\bibitem{male00} Y. Mal\'{e}cot, C. Auriault, H. Kahalerras, Y. Gagne,
  O.   Chanal,  B.  Chabaud,  and  B.   Castaing, \bibinfo{title}  ``A
  statistical  estimator of turbulence  intermittency in  physical and
  numerical  experiments'',   \bibinfo{journal}{Eur.   Phys.   J.   B}
  \textbf{\bibinfo{volume}{16}},                   \bibinfo{pages}{549}
  (\bibinfo{year}{2000}).

\bibitem{venu06a} V. Venugopal, S.  G. Roux, E. Foufoula-Georgiou, and
  A.    Arn\'{e}odo,  \bibinfo{title}   ``Scaling  behavior   of  high
  resolution  temporal rainfall: {N}ew  insights from  a wavelet-based
  cumulant    analysis'',     \bibinfo{journal}{Phys.     Lett.     A}
  \textbf{\bibinfo{volume}{348}},                  \bibinfo{pages}{335}
  (\bibinfo{year}{2006a}).

\bibitem{venu06b} V. Venugopal, S.  G. Roux, E. Foufoula-Georgiou, and
  A.   Arn\'{e}odo,  \bibinfo{title}  ``Revisiting multifractality  of
  high-resolution temporal rainfall using a wavelet-based formalism'',
  \bibinfo{journal}{Wat.  Resour. Res.} \textbf{\bibinfo{volume}{42}},
  \bibinfo{pages}{DOI:10.1029/2005WR004489} (\bibinfo{year}{2006b}).

\bibitem{muzy93}  J.   F.   Muzy,  E.   Bacry,  and  A.   Arn\'{e}odo,
  \bibinfo{title}  ``Multifractal formalism  for fractal  signals: The
  structure-function    approach    versus    the    wavelet-transform
  modulus-maxima    method'',   \bibinfo{journal}{Phys.     Rev.    E}
  \textbf{\bibinfo{volume}{47}},                   \bibinfo{pages}{875}
  (\bibinfo{year}{1993}).

\bibitem{arne95}  A.   Arn\'{e}odo,  E.   Bacry,  and  J.   F.   Muzy,
  \bibinfo{title}  ``The  thermodynamics  of fractals  revisited  with
  wavelets'',               \bibinfo{journal}{Physica               A}
  \textbf{\bibinfo{volume}{213}},                  \bibinfo{pages}{232}
  (\bibinfo{year}{1995}).

\bibitem{muzy91}  J.   F.   Muzy,  E.   Bacry,  and  A.   Arn\'{e}odo,
  \bibinfo{title} ``Wavelets  and multifractal formalism  for singular
  signals:  Application to turbulence  data'', \bibinfo{journal}{Phys.
  Rev.   Lett.}  \textbf{\bibinfo{volume}{67}},  \bibinfo{pages}{3515}
  (\bibinfo{year}{1991}).

\bibitem{verg93}  M.   Vergassola, R.   Benzi,  L.   Biferale, and  D.
  Pisarenko,   \bibinfo{title}  ``Wavelet   analysis  of   a  Gaussian
  Kolmogorov     signal'',     \bibinfo{journal}{J.      Phys.      A}
  \textbf{\bibinfo{volume}{26}},                  \bibinfo{pages}{6093}
  (\bibinfo{year}{1993}).

\bibitem{thei92} J. Theiler, S. Eubank, A. Longtin, B. Galdrikian, and
  J.  D.  Farmer, \bibinfo{title}  ``Testing for nonlinearity  in time
  series: the method of surrogate data'', \bibinfo{journal}{Physica D}
  \textbf{\bibinfo{volume}{58}},                    \bibinfo{pages}{77}
  (\bibinfo{year}{1992}).

\bibitem{kant97} H. Kantz and  T. Schriber, {\it Nonlinear Time Series
Analysis} (Cambridge University Press, Cambridge, UK, 1997), p. 320.

\bibitem{basu02}  S. Basu  and  E. Foufoula-Georgiou,  \bibinfo{title}
  ``Detection of nonlinearity and  chaoticity in time series using the
  transportation distance function'', \bibinfo{journal}{Phys. Lett. A}
  \textbf{\bibinfo{volume}{301}},                  \bibinfo{pages}{413}
  (\bibinfo{year}{2002}).

\bibitem{vene06a} V. Venema, F.  Ament, and C. Simmer, \bibinfo{title}
  ``A  Stochastic  Iterative   Amplitude  Adjusted  Fourier  Transform
  Algorithm with improved accuracy'', \bibinfo{journal}{Nonlin.  Proc.
  Geophys.}     \textbf{\bibinfo{volume}{13}},    \bibinfo{pages}{247}
  (\bibinfo{year}{2006}).

\bibitem{venu05} V.   Venugopal, S.  Basu,  and E.  Foufoula-Georgiou,
  \bibinfo{title} ``A new  metric for comparing precipitation patterns
  with  an application  to ensemble  forecasts'', \bibinfo{journal}{J.
  Geophys.            Res.}            \textbf{\bibinfo{volume}{110}},
  \bibinfo{pages}{DOI:10.1029/2004JD005395} (\bibinfo{year}{2005}).

\bibitem{vene06b} V.  Venema, S. Meyer, S. G.  Garc\'{i}a, A. Kniffka,
  C. Simmer, S.  Crewell, U.  L\"{o}hnert, T. Trautmann, and A. Macke,
  \bibinfo{title}   ``Surrogate  cloud   fields  generated   with  the
  Iterative   Amplitude   Adapted   Fourier   Transform   algorithm'',
  \bibinfo{journal}{Tellus}             \textbf{\bibinfo{volume}{58A}},
  \bibinfo{pages}{104} (\bibinfo{year}{2006}).

\bibitem{niko01} V. Nikora, D. Goring, and R. Camussi, \bibinfo{title}
  ``Intermittency  and interrelationships  between  turbulence scaling
  exponents:   Phase-randomization   tests'',  \bibinfo{journal}{Phys.
  Fluids}     \textbf{\bibinfo{volume}{13}},     \bibinfo{pages}{1404}
  (\bibinfo{year}{2001}).
  
\bibitem{pahl01} M.  Pahlow, M.  B.  Parlange, and  F. Port\'{e}-Agel,
  \bibinfo{title}  ``On  Monin Obukhov  similarity  in  the  stable
  atmospheric   boundary   layer'',   \bibinfo{journal}{Boundary-Layer
  Meteorol.}     \textbf{\bibinfo{volume}{99}},   \bibinfo{pages}{225}
  (\bibinfo{year}{2001}).

\bibitem{basu06} S.  Basu, F.  Port\'{e}-Agel,  E.  Foufoula-Georgiou,
  J.-F.  Vinuesa,  M.  Pahlow, \bibinfo{title}  ``Revisiting the local
  scaling hypothesis  in stably stratified  atmospheric boundary-layer
  turbulence: An integration of field and laboratory measurements with
  large-eddy      simulations'',      \bibinfo{journal}{Boundary-Layer
  Meteorol.}    \textbf{\bibinfo{volume}{119}},   \bibinfo{pages}{473}
  (\bibinfo{year}{2006}).
  
\bibitem{arne00}  A.  Arn\'{e}odo,  N.   Decoster, and  S.  G.   Roux,
  \bibinfo{title}  ``A  wavelet-based  method for  multifractal  image
  analysis.   I. Methodology  and test  applications on  isotropic and
  anisotropic  random rough surfaces'',  \bibinfo{journal}{Eur.  Phys.
  J.     B}     \textbf{\bibinfo{volume}{15}},    \bibinfo{pages}{567}
  (\bibinfo{year}{2000}).

\bibitem{aiva02} K. G. Aivalis, K. R. Sreenivasan, J. C. Klewicki, and
  C.  A.,  Biltoft, \bibinfo{title} ``Temperature  structure functions
  for     air     flow     over    moderately     heated     ground'',
  \bibinfo{journal}{Phys.     Fluids}   \textbf{\bibinfo{volume}{14}},
  \bibinfo{pages}{2439} (\bibinfo{year}{2002}).
  
\bibitem{shi05}  B.  Shi,  B.  Vidakovic,  G.  G.   Katul, and  J.  D.
  Albertson,  \bibinfo{title} ``Assessing  the effects  of atmospheric
  stability on  the fine structure  of surface layer  turbulence using
  local  and global  multiscale  approaches'', \bibinfo{journal}{Phys.
  Fluids}    \textbf{\bibinfo{volume}{17}},    \bibinfo{pages}{055104}
  (\bibinfo{year}{2005}).

\bibitem{poul02}   G.   S.   Poulos,   W.  Blumen,   D.   C.   Fritts,
J. K. Lundquist,  J. Sun, S. P. Burns, C. Nappo,  R. Banta, R. Newsom,
J. Cuxart, E. Terradellas,  B. Balsley, and M. Jensen, \bibinfo{title}
``CASES-99:  A  Comprehensive Investigation  of  the Stable  Nocturnal
Boundary   Layer'',  \bibinfo{journal}{Bull.  Amer.   Meteorol.  Soc.}
\textbf{\bibinfo{volume}{83}},                     \bibinfo{pages}{555}
(\bibinfo{year}{2002}).

\end{thebibliography}
\end{document}